\documentclass[12pt]{article}
\usepackage{enumerate}
\usepackage[margin=1.25in]{geometry}                
\usepackage{comment}
\usepackage{authblk}
\usepackage[normalem]{ulem}
\usepackage{caption}
\usepackage{subcaption}
\usepackage{graphicx}
\usepackage{float}
\usepackage{amssymb}
\usepackage[english]{babel}
\usepackage{epstopdf} 
\usepackage{amsmath}
    \numberwithin{equation}{section}
\usepackage{hyperref} 
\newcommand{\appendixpage}

\hypersetup{
    pdfstartview={FitH},    
    colorlinks=true,       
    linkcolor=black,          
    citecolor=black,        
    filecolor=blue,      
    urlcolor=black          
}

\def\expec#1{\langle #1 \rangle}

\newcommand{\cB}{{\mathcal{B}}}

\newcommand{\cO}{{\mathcal{O}}}

\newcommand{\cS}{{\mathcal{S}}}

\def\ren{\ensuremath{_{\mbox{\tiny ren}}}}
\def\bare{\ensuremath{_{\mbox{\tiny bare}}}}
\def\sub{\ensuremath{_{\mbox{\tiny sub}}}}
\def\reg{\ensuremath{_{\mbox{\tiny reg}}}}
\def\os{\ensuremath{^{\mbox{\tiny on-shell}}}}
\def\ct{\ensuremath{_{\mbox{\tiny ct}}}}
\def\intr{\ensuremath{^{\mbox{\tiny int}}}_{\mbox{\tiny ct}}}
\def\ext{\ensuremath{^{\mbox{\tiny ext}}}_{\mbox{\tiny ct}}}
\def\hyb{\ensuremath{^{\mbox{\tiny hyb}}}_{\mbox{\tiny ct}}}
\def\sol{\ensuremath{_{\mbox{\tiny sol}}}}

\begin{document}

\definecolor{orange}{rgb}{0.9,0.45,0}
\definecolor{blue}{rgb}{0.0,0.0,1.0}
\definecolor{verde}{rgb}{0.0, 0.7, 0.5}
\newcommand{\ag}[1]{{\textcolor{verde}{#1}}}
\newcommand{\agcomment}[1]{{\textcolor{verde}{[AG: #1]}}}
\newcommand{\spl}[1]{{\textcolor{orange}{[SPL: #1]}}}
\newcommand{\ija}[1]{{\textcolor{red}{[IJA: #1]}}}
\newcommand{\ga}[1]{{\textcolor{blue}{[GA: #1]}}}

\title{\bf{Extrinsic Holographic Renormalization \\ 
for a Scalar Field}}
\author[1]{\thanks{\href{mailto: georgios.anastasiou@uai.cl}{georgios.anastasiou@uai.cl}}Giorgos Anastasiou}
\author[2]{\thanks{\href{mailto: araya.quezada.ignacio@gmail.com}{araya.quezada.ignacio@gmail.com}}Ignacio J. Araya}
\author[3]{\thanks{\href{mailto: daniel.avila@correo.nucleares.unam.mx}{daniel.avila@correo.nucleares.unam.mx}}Daniel Ávila}
\author[3]{\\ \thanks{\href{mailto: alberto@nucleares.unam.mx}{alberto@nucleares.unam.mx}}Alberto Güijosa}
\author[3]{\thanks{\href{mailto:sergio.patino@correo.nucleares.unam.mx}{sergio.patino@correo.nucleares.unam.mx}}Sergio Patiño-López}

\affil[1]{\small\it{Universidad Adolfo Ibáñez, Facultad de Artes Liberales, Departamento de Ciencias,
Av.~Diagonal Las Torres 2640, Peñalolen, Chile}}
\affil[2]{\small{\it{Universidad Andres Bello, Departamento de Ciencias F\'isicas, Facultad de Ciencias Exactas, Sazi\'e 2212, Piso 7, Santiago, Chile}}}
\affil[3]{\it{Departamento de Física de Altas Energías, Instituto de Ciencias Nucleares, Universidad~Nacional Autónoma de México, Apdo.~Postal 70-543, CdMx 04510, México}}

\date{}                                           

\maketitle

\begin{abstract}
In \hfill the \hfill context \hfill of \hfill the \hfill holographic \hfill correspondence, we \hfill introduce \hfill a \hfill purely  \\
\emph{extrinsic} renormalization prescription, exemplified with the case of a minimally-coupled scalar field in AdS space.  The counterterms depend only on the field and its radial derivatives. This would seem to conflict with the Dirichlet variational principle, but we show that consistency follows from the fact  that the asymptotic structure of asymptotically locally AdS spacetimes requires not only the leading, but also all of the subleading non-normalizable modes to be fixed as a boundary condition. Crucially, as seen from a path integral definition of the bulk partition function involved in the standard GKPW formula, this condition is valid  away from the saddle. 
We find that the extrinsic renormalization prescription is maximally efficient when the scalar field is massless, which is suggestive of a connection with the Kounterterm method for renormalization of pure gravity.
\\

\textbf{Keywords}: AdS/CFT, holographic renormalization, scalar field   

\end{abstract}

\newpage
\tableofcontents

\section{Introduction}\label{introsec}

The anti-de Sitter (AdS)/Conformal Field Theory (CFT) correspondence~\cite{Maldacena:1997re,Gubser:1998bc,Witten:1998qj} resides in the core of the gauge/gravity duality, which claims the equivalence between certain gravitational theories in $d+1$ dimensions and certain field theories in one dimension lower. In particular, `bulk' actions on asymptotically locally AdS (alAdS) spacetimes can be used as a tool to extract `boundary' information regarding correlation functions 
of a certain class of CFTs, dubbed holographic. In more detail, the GKPW prescription \cite{Gubser:1998bc,Witten:1998qj} equates the bulk and boundary partition functions, implying that in the limit of large central charge, the on-shell
action on AdS is interpreted as the saddle-point approximation to the generating functional for connected correlators in the CFT. The (rescaled) value of each bulk field is  held fixed at the conformal boundary of AdS, and corresponds to the external source (equivalently, the spacetime-dependent coupling) for the dual CFT operator. As such, two related requirements are imposed on a well-defined AdS action. Firstly, it must be finite, and secondly, its generic on-shell variation must be Dirichlet with respect to the holographic sources, which are the aforementioned boundary field values.\footnote{For fields of mass within a restricted range \cite{Breitenlohner:1982bm,Breitenlohner:1982jf}, there exists the alternative of imposing Neumann or mixed boundary conditions, corresponding to studying the CFT at a different fixed point or a multi-trace deformation thereof \cite{Klebanov:1999tb,Minces:1999eg,Berkooz:2002ug,Sever:2002fk,Papadimitriou:2007sj}.} This ensures that the functional derivative of the action is a total functional derivative on the sources, which is a required property of the CFT partition function.

As it is well known, the metric of alAdS spacetimes is characterized by the presence of a second order pole in the radial direction. This induces infinities asymptotically, leading the bare on-shell action to diverge. By virtue of the UV-IR connection \cite{Susskind:1998dq,Peet:1998wn}, these IR divergences of the gravity side get mapped into the usual UV divergences of the CFT partition function. As a consequence, the definition of the renormalized gravitational action requires the addition of terms that eliminate the divergences without modifying the equations of motion. They are called counterterms, in analogy with the usual field-theoretic renormalization procedure. 

A systematic procedure for renormalizing the bulk action was developed in \cite{Skenderis:1999nb,deHaro:2000vlm,Skenderis:2002wp}. This is based on the introduction of a regulator at a finite radial distance, which, by use of a near-boundary (FG)
expansion of the metric \cite{Fefferman:1985} and other bulk fields, allows to identify the divergent contributions to the action.\footnote{These arise both from the bulk terms and from the familiar Gibbons-Hawking-York (GHY) boundary term, which would suffice to ensure a well-defined Dirichlet variational principle for the gravitational action if one were dealing with an actual boundary at finite radial distance, instead of having a radial cutoff that needs to be removed to revert to the conformal boundary.} 
One then cancels these divergences by adding boundary terms constructed out of the induced fields on the boundary, in order to preserve the Dirichlet variational principle.
This standard formulation of holographic renormalization provides a series of counterterms involving a successively larger number of \emph{intrinsic} derivatives of the bulk fields (i.e., derivatives along the directions that are tangential to the regulated boundary).

The counterterms not only 
lead to a finite action, but also provide a well-posed variational problem \cite{Papadimitriou:2010as}. This insight, together with the 
known relation between 
radial (PBH) diffeomorphisms in the bulk
and Weyl transformations of the boundary metric \cite{Imbimbo:1999bj},
was exploited in \cite{Papadimitriou:2003is,Papadimitriou:2004ap} (see also \cite{Kraus:1999di}) to provide a Hamiltonian reformulation\footnote{Hamiltonian, that is, with the radial direction playing the role of time.} of holographic renormalization in terms of  eigenfunctions of the dilatation operator at the conformal boundary, which
coincides with the radial derivative at leading order.
It was shown that finding the renormalized action amounts to adding intrinsic correction terms which ensure that the variation of the action is dilatation-invariant. This prescription is both elegant and efficient, and it makes the conformal invariance of the boundary theory manifest.

For the special case of pure gravity, an interesting alternative to standard and Hamiltonian holographic renormalization, known as the \textit{Kounterterm} method, was developed in~\cite{Olea:2005gb,Olea:2006vd}. In this approach, there is a single boundary form\footnote{Used instead of the GHY term.} that accomplishes the dual role of canceling divergences and imposing the Dirichlet variational principle with respect to the holographic source of the CFT stress tensor encoded in the boundary metric.\footnote{The holographic source is given by the leading order coefficient $h_{(0)ij}$ of the FG expansion of the boundary metric, defined below in (\ref{MetricGauss})-(\ref{FGh}).} The form in question has a topological origin, and its definition is different depending on whether the bulk manifold has even or odd dimension. For even dimensions, it is given by the 
Chern form associated to the Euler density in the bulk \cite{Olea:2005gb}. For odd dimensions, it corresponds to the contact term of the Chern-Simons transgression form associated to the AdS group~\cite{Olea:2006vd}. Another attractive feature of this method is that the same boundary form serves to renormalize different theories of gravity, up to the adjustment of a single overall numerical coefficient \cite{Kofinas:2007ns,Giribet:2018hck}. A limitation of the Kounterterm approach is that, above four dimensions, it does not apply to generic alAdS geometries, but only to the subclass where the boundary Weyl tensor vanishes \cite{Anastasiou:2020zwc}. Recently, this limitation has been overcome by an extension known as conformal renormalization \cite{Anastasiou:2021tlv,Anastasiou:2023oro, Anastasiou:2022wjq}.  

A prominent characteristic of the Kounterterm method is that, unlike holographic renormalization, it involves
\emph{extrinsic} derivatives of the metric. 
Indeed, the boundary form is constructed with certain products of the intrinsic Riemann tensor $\mathcal{R}$ and the extrinsic curvature $K$ of the regulated boundary, the latter defined, as usual, by the Lie derivative of the induced metric along the holographic radial coordinate. The conspicuous presence of $K$ is the reason why one speaks of Kounterterms instead of counterterms. A priori, this seems incompatible with the desired Dirichlet condition for the boundary metric, but detailed arguments were provided to prove that, due to the special asymptotics of alAdS spacetimes, there is in fact no such incompatibility \cite{Miskovic:2008ck,Miskovic:2010ui,Anastasiou:2019ldc}.

Knowing that Kounterterms successfully renormalize the metric employing extrinsic derivatives, one is naturally led to ask  whether an analogous renormalization procedure might be feasible for the matter sector of the theory. In the present paper we will answer this question in the affirmative, by considering, as the simplest possible example, a scalar field minimally coupled to AdS gravity, neglecting the backreaction of the scalar on the metric. We construct purely extrinsic counterterms for the scalar action, considering only the scalar field and its derivatives with respect to the holographic radial coordinate. We show that this procedure leads to a consistent description which can be developed independently of the standard holographic renormalization approach, as it does not need its input.

Due to this independence, one potential concern is that our procedure might turn out to be inequivalent to the standard approach. This would be disallowed in the context of holography, because the above-described contact with the CFT partition function demands that the counterterms be fully expressible in \emph{intrinsic} form. This concern can be dispelled by the following argument. The central requirement for any set of purported counterterms is to cancel the divergences of the bare bulk action, so any two choices of counterterms can at most differ at finite level. If the finite piece is scheme-dependent, then such a difference would be inconsequential. If the correct finite piece is uniquely defined, then it is fixed by Ward identities, which is to say, by symmetries. Altogether, then, at the level of the action itself, any alternative choice of counterterms is guaranteed to lead to the same result as the standard, intrinsic choice if it correctly cancels all divergences and respects the same symmetries.

Knowing that such alternative choices are bound to lead to the same result for the on-shell renormalized action, we must additionally verify that their \emph{variations} with respect to the CFT sources coincide. More specifically, for anyone familiar with the standard approach to holographic renormalization, the main unease with extrinsic counterterms is that they would seem to conflict with the Dirichlet variational principle, which would make them unviable. Mindful consideration is thus necessary to determine whether this conflict is avoided in the case of a scalar field, just like it turned out to be avoided for pure gravity in the Kounterterm approach to renormalization.

The paper is organized as follows. In Section~\ref{Asymptotic}, we argue that, unlike in the familiar situation where one has a literal boundary at finite distance, on an alAdS background the conformal nature of the boundary implies that all field configurations in the path integral must be \emph{asymptotically on shell}, up to normalizable order. 
As we explain, the asymptotic validity of the equation of motion implies
that, \emph{even for the variation of the field}, the equation of motion allows one to interchange radial derivatives with tangential derivatives at the conformal boundary, and this in turn ensures that, contrary to appearances, boundary terms with radial derivatives present no challenge to the Dirichlet variational principle. This insight explains and extends the corresponding findings of \cite{Miskovic:2008ck,Miskovic:2010ui,Anastasiou:2019ldc} in the context of Kounterterms, and is crucial for the main goal of the present paper. As we also point out in Section~\ref{Asymptotic}, there is in fact an infinite number of different but equivalent versions of the counterterms, and we choose to focus on the purely extrinsic version for reasons explained there. 

In an attempt to make this paper moderately self-contained, Section~\ref{Intrinsic} presents a brief summary of intrinsic holographic renormalization, including both the standard and the more modern Hamiltonian formulations. 

In Section~\ref{masslesssec} we present our extrinsic renormalization prescription in the concrete example of a free massless scalar field on AdS, showing how to construct counterterms that depend only on the field and its radial derivatives at the conformal boundary. We frame the prescription in terms of the eigenfunctions of the radial derivative operator, which can be directly related to the FG expansion coefficients.  In Sections~\ref{Extrinsicgeneric} and~\ref{Extrinsiclog}, we extend the method to the case where the scalar field is massive, with the bulk spacetime dimension being respectively even or odd. 

Finally, in Section~\ref{conclusiones} we summarize our  conclusions and give an overview of possible extensions of our results  in future work. 

Two appendices provide further detail on a couple of aspects of our prescription: Appendix~\ref{translationapp} contemplates the passage from intrinsic to extrinsic counterterms via direct translation, and Appendix~\ref{wardapp} reviews the steps involved in verifying that the Ward identity for diffeomorphisms is satisfied.

\section{Asymptotic Restriction on the Variational Principle}\label{Asymptotic}
 
 In a classical field theory, the equations of motion and boundary conditions are on an equal footing: they both follow from the variational principle, and are both satisfied by the allowed field configurations.
 At the quantum level, a clear disparity emerges. The boundary conditions continue to hold firmly: they define the set $\cB$ 
 of field configurations  that we sum over in the path integral. The equations of motion, on the other hand, are not enforced within the integrand. They are only satisfied on average (in the sense of Schwinger-Dyson equations), or in some cases, within a properly justified saddle-point approximation.

 In \hfill the \hfill AdS/CFT \hfill context, \hfill the \hfill bulk \hfill description has \hfill a \emph{conformal} \hfill boundary \hfill at\\
 $r\to\infty$, and the holographically renormalized GKPW path integral\footnote{For concreteness, throughout the paper we refer to the Euclidean partition function, as in \cite{Gubser:1998bc,Witten:1998qj}, but our discussion and results can equally well be framed in terms of the Lorentzian formulation of \cite{Balasubramanian:1998sn,Skenderis:2008dh,Skenderis:2008dg}. }
\begin{equation}
Z[\phi_{(0)}]=\int_{\Phi\in\cB[\phi_{(0)}]}
\mathcal{D}\Phi\,\text{exp}(-S\ren)~,
\label{partitionfcn}
\end{equation}
automatically discards all field configurations with divergent renormalized action 
$S\ren\equiv S\bare+S\ct~.$ Since the counterterm action is constructed to cancel divergences of the \emph{on-shell} $S\bare$, this implies that the set $\cB[\phi_{(0)}]$ consists of all field configurations that match asymptotically  onto a solution $\Phi\sol[\phi_{(0)}]$ of the equations of motion whose leading (non-normalizable) behavior is specified by $\phi_{(0)}(x)$. And, importantly, \emph{this matching must hold to all orders in the radial expansion that would give rise to divergences}.

We will use the notation $\Phi\simeq\Phi\sol[\phi_{(0)}]$ to express this specific sense of asymptotic matching, which is a feature of the GKPW prescription that is not usually highlighted.\footnote{Reference \cite{Papadimitriou:2010as} comes close to this notion, but it does not spell out the need for the matching to extend beyond the leading-order coefficient $\phi_{(0)}$. The idea is implicit in \cite{Balasubramanian:1998sn}, which pre-dates the use of counterterms for holographic renormalization.} 
The $\simeq$ equivalence relation is well-defined due to the fact that distinct choices of $\Phi\sol[\phi_{(0)}]$ differ from one another only at normalizable order, and therefore give rise to identical divergences in $S\bare.$

The normalizable vs.~non-normalizable distinction is most often discussed in the context of \emph{on-shell} bulk field configurations, where (for second-order systems) one has a single mode of each type \cite{Gubser:1998bc,Witten:1998qj}.  The familiar statement then is that the non-normalizable mode $\phi_{(0)}(x)$ ought to be fixed classically, as a boundary condition, while the normalizable, fluctuating mode $\phi_{(\Delta)}(x)$ is subjected to quantization.\footnote{The precise meaning of the modes $\phi_{(k)}(x)$ will be reviewed in the next section. See (\ref{FGPhi}).}\textsuperscript{,}\footnote{For fields with sufficiently low masses, the distinction between normalizable and non-normalizable is ambiguous, and one can freely choose to assign each of the two independent modes to one category or the other \cite{Breitenlohner:1982bm,Breitenlohner:1982jf,Klebanov:1999tb,Berkooz:2002ug,Marolf:2006nd,Compere:2008us}.} The simple point that we wish to emphasize here is that, within the path integral (\ref{partitionfcn}), even after fixing $\phi_{(0)}(x)$ there exist several other non-normalizable modes 
$\phi_{(0<k<\Delta)}(x)$ that would be \emph{a priori} undetermined, but their values too must be considered to be fully specified as boundary conditions: they are prescribed to be on shell, and are consequently determined by $\phi_{(0)}(x)$.\footnote{For simplicity, we are phrasing our discussion here only in terms of the modes appearing in the familiar FG radial expansion (\ref{FGPhi}), but a priori there could also exist off-shell modes with unfamiliar radial dependence (e.g., $r\,e^{-(d-\Delta)r}$), and our argument implies that the coefficients of any such modes that are non-normalizable ought to be set to zero as a boundary condition.}

Our final observation in this section is that the variational principle is by design considered only within the set  $\cB[\phi_{(0)}]=\{\Phi~|~\Phi\simeq\Phi\sol[\phi_{(0)}]\}$ of configurations included in the path integral,\footnote{Put another way, the boundary conditions prescribed for any path integral must always be compatible with the variational principle derived from the corresponding action \cite{BrinkHenneaux}.} so it must respect the asymptotic matching. We conclude then that, contrary to the familiar case with a true boundary, where the field variation $\delta\Phi$ obeys the boundary conditions but is not restricted by the equation of motion, in the holographic setting the conformal nature of the boundary implies that $\delta\Phi$ \emph{must be asymptotically on-shell}, in the specific sense described above. 
In terms of the radial expansion, this fixes the radial dependence of the variation up to normalizable order, and implies that the successive coefficients in the expansion are related in precisely the same fashion as for $\Phi$ itself (see (\ref{EOMSolution}) below).  

For our purposes in the present paper, this conclusion is important for the following reason. By design, the standard counterterms $S\ct$ employed for holographic renormalization are boundary terms involving only intrinsic derivatives. As long as we restrict ourselves to considering on-shell field configurations, we can use the field equation of motion to rewrite $S\ct$ trading all intrinsic for extrinsic derivatives. But these two alternative presentations, $S\intr$ and $S\ext$, would a priori \emph{not} be expected to agree at the level of the variational principle. Since $\delta\Phi$ is generally off-shell, and extrinsic derivatives cannot be integrated by parts along the boundary, it is only $S\intr$ that is manifestly compatible with the usual phrasing of the GKPW prescription, in terms of the Dirichlet boundary condition  $\delta\phi_{(0)}=0$. This is why the papers that developed extrinsic renormalization for the metric made a point of verifying that, contrary to appearances, the Kounterterms do not in fact interfere with the demand that $\delta h_{(0)ij}=0$ \cite{Olea:2005gb,Olea:2006vd,Anastasiou:2020zwc,Miskovic:2008ck,Miskovic:2010ui,Anastasiou:2019ldc,Aros:1999kt,Mora:2004rx,Miskovic:2006tm}.
However, in the previous paragraph we have learned that, in holography, the conformal nature of the boundary implies that $\delta\Phi$ itself ought to be asymptotically on-shell, allowing us to switch from intrinsic to extrinsic derivatives without any concern about spoiling the Dirichlet variational principle.  This validates our pursuit of an extrinsic method of renormalization in the remainder of the paper.

Notice that, if we visualize $S\ext$ as  being obtained from $S\intr$ by using the equation of motion to remove all of the intrinsic derivatives, then we encounter the alternative possibility of removing only some of said derivatives. This would lead to an intermediate version of the counterterms, which makes use of both intrinsic and extrinsic derivatives. Such `hybrid' counterterms provide a more direct analog of the metric Kounterterms, and are explored in Appendix~\ref{translationapp}. More generally, one could replace each intrinsic derivative with some linear combination of intrinsic and extrinsic derivatives, so there is in fact an infinite-dimensional set of different but equivalent presentations of the counterterms. We know of their equivalence because we are assuming in this paragraph that they are all just different rewritings of $S\intr$.\footnote{We are referring here to equivalence as far as the bulk-with-conformal-boundary variational problem is concerned, which is the relevant setup for examining renormalized correlators in a holographic quantum field theory at arbitrarily large energies, all the way up to the UV fixed point. A different question is how the system evolves under the holographic Wilsonian renormalization group, which amounts to moving the alAdS boundary inward, to a finite radial location \cite{Heemskerk:2010hk,Faulkner:2010jy,Balasubramanian:2012hb}. The intrinsic presentation of the counterterms is advantageous for that analysis, which is not the subject of the present paper.} And even if we were led to any of them by some independent procedure, equivalence to the intrinsic counterterms would be guaranteed by the argument of Section~\ref{introsec}. 

In Sections~\ref{masslesssec}, \ref{Extrinsicgeneric} and \ref{Extrinsiclog}, we will prefer to pursue a purely extrinsic choice of counterterms, $S\ext$, for three reasons. First, it is of course conceptually simpler to employ derivatives of only a single type. Second, $S\ext$ is diametrically opposed to the standard $S\intr$, so without the insight gained from this section, it would have seemed maximally problematic for compatibility with a Dirichlet boundary condition. Third and most important, we will find that $S\ext$ can be obtained through a completely self-contained procedure, not requiring previous knowledge of $S\intr$, and moreover, the procedure in question is highly efficient in the case where the field is massless.

\section{Brief Review of the Intrinsic Counterterms \label{Intrinsic}}
In this section we recall how the intrinsic counterterms are constructed, working in the simplest possible setting: a free scalar field $\Phi$ propagating on a fixed Euclidean asymptotically AdS$_{d+1}$ background. We choose units such that the curvature radius is $\ell=1$, and work in Gaussian normal coordinates, such that the line element takes the form
\begin{equation}
ds^{2}=g_{\mu\nu}dx^{\mu}dx^{\nu}=dr^{2}+h_{ij}(r,x)dx^{i}dx^{j},
\label{MetricGauss}
\end{equation}
with the boundary of the spacetime being located at $r\to\infty$. Here $h_{ij}$ denotes the induced metric on constant-$r$ hypersurfaces $\Sigma_{r}$. In the case of pure AdS, one can choose the radial foliation such that $h_{ij}=e^{2r}\delta _{ij}\,.$ For asymptotically locally AdS (alAdS) spacetimes, the induced metric can be written out in terms of 
the well-known Fefferman-Graham (FG) expansion \cite{Skenderis:1999nb,deHaro:2000vlm,Fefferman:1985}
\begin{align}
h(r,x)&=e^{2r}\left(h_{( 0)}(x)+e^{-2r}h _{(2)}(x)+\cdots
+e^{-d r}\left(h_{(d)}(x)+r \tilde{h}_{(d)}(x)\right)+\cdots\right)~,
\label{FGh}
\end{align}
where the spacetime subindices $_{ij}$ have been left implicit. The term involving $\tilde{h}_{(d)}$ is present for even $d$, and is the embodiment of the metric contribution to the Weyl anomaly in the CFT \cite{Witten:1998qj,Henningson:1998gx,Henningson:1998ey}. Its linear dependence on $r$ amounts to a logarithmic dependence in the most-often-used radial coordinate $z\equiv e^{-r}$. 
The Einstein equation determines in terms of the non-normalizable mode
$h_{(0)}$ the functions $h_{(k<d)}$ and $\tilde{h}_{(d)}$, as well as the trace and covariant divergence of $h_{(d)}$.
For spacetimes that are asymptotically AdS (aAdS), which will be the focus of the present paper, 
$h_{(0)ij}=\delta _{ij}$ and 
$h_{(k<d)}=\tilde{h}_{(d)}=0$, so only the normalizable mode $h_{(d)}$ is kept arbitrary.

The action of the scalar field is
\begin{equation}
S\bare=\frac{1}{2}\int d^{d+1}x \sqrt{g}(\partial^{\mu}\Phi\partial_{\mu}\Phi+m^{2}\Phi^{2})~,
\label{BulkAction}
\end{equation}
where the mass $m$ is related to the scaling dimension $\Delta$ of the dual operator $\cO$ in the CFT by $m^{2}=\Delta(\Delta-d)$. Generically, one is interested in the largest root of this equation,
\begin{equation}\label{delta}
    \Delta= \frac{d}{2}+\sqrt{\frac{d^2}{4}+m^2}~.
\end{equation}
The equation of motion that follows from (\ref{BulkAction}) is of course 
\begin{equation}
\Box_{g}\Phi-m^{2}\Phi=0~,
\label{EOM}
\end{equation}
where $\Box_{g}$ denotes the Laplacian with respect to the metric $g$, $\Box_{g}\Phi=\frac{1}{\sqrt{g}}\partial_{\mu}(\sqrt{g}g^{\mu\nu}\partial_{\nu}\Phi)$.

We are interested in the UV divergences of the action (\ref{BulkAction}) when it is evaluated on shell, hence we solve (\ref{EOM}) by series expansion near the boundary. The details depend on the mass $m$, or equivalently, on the scaling dimension $\Delta$. One crucial datum is the discrete variable
\begin{equation}
    n\equiv \lfloor\,\Delta-\frac{d}{2}\, \rfloor~,
    \label{ndef}
\end{equation}
i.e., the non-negative integer such that the scaling dimension (\ref{delta}) lies in the window 
\begin{equation}\label{window}
\frac{d}{2}+n\le \Delta<\frac{d}{2}+n +1~.
\end{equation}
Within the GKPW prescription, one is most often interested in correlators of relevant or marginal operators, for which finite source terms can be turned on. This requires $\Delta\le d$, which translates into the upper bound $n\le \lfloor\,d/2\, \rfloor$. Holographic correlators for irrelevant operators have been discussed in \cite{vanRees:2011fr}.

In terms of $\Delta$ and $n$, the radial expansion takes the FG form \cite{deHaro:2000vlm,Fefferman:1985}
\begin{align}
\Phi&=e^{-\left (d-\Delta\right )r}\phi _{\left ( 0 \right )}+e^{-\left (d-\Delta+2\right)r}\phi _{\left (2 \right )}+\,\cdots\, +e^{-\left (d-\Delta+2n\right)r}\phi_{\left (2n\right )}\nonumber\\
{}&\quad
+\delta_{\Delta,\frac{d}{2}+n}\,e^{-\Delta r}
\left(-\phi_{(2n)}
+r\tilde{\phi}_{(\Delta)}\right)
+e^{-\Delta r}\phi_{(\Delta)}+\,\cdots~.
\label{FGPhi}
\end{align} 
 As indicated by the Kronecker delta, a special situation arises when
 $\Delta=\frac{d}{2}+n$, i.e., when the left inequality in (\ref{window}) is saturated: in that case the $\phi_{(2n)}$ term, which would have had the same radial dependence as $\phi_{(\Delta)}$, is absent, and its place is taken by the $\tilde{\phi}_{(\Delta)}$ term, which has an additional, linear-in-$r$ dependence, which is logarithmic with respect to $z\equiv e^{-r}$. This term embodies the matter contribution to the Weyl anomaly in the CFT \cite{deHaro:2000vlm,Petkou:1999fv}.
  
  Solving the equation of motion \eqref{EOM} order by order in $e^r$, one can determine all of the coefficients $\phi_{(k)}$ in the first line of (\ref{FGPhi}) (i.e., those with $k\le 2n$) as local functions of $\phi_{(0)}$,
\begin{equation}
\phi_{(2k)}=\frac{1}{2k(2\Delta-d-2k)}\Box_{\left(0\right)}\phi_{(2k-2)}~.
\label{EOMSolution}
\end{equation}
In the special case $\Delta=\frac{d}{2}+n$, the equation of motion determines not $\phi_{(2n)}$, but
\begin{equation}
\tilde{\phi}_{(\Delta)}=\frac{1}{2^{2n-1}\Gamma(n)\Gamma(n+1)}\left(\Box_{\left(0\right)}\right)^{n}\phi_{(0)}~.
\end{equation}
Since $h_{ij}$ is (to leading order in $r$) a rescaling of $h_{\left(0\right) ij}$, the respective Laplacian operators are related by
\begin{equation}
\Box_{\left(0\right)} = e^{2r} \Box_{h} \,.
\end{equation}
For all values of $\Delta$, the two independent solutions to the equation of motion are parametrized by $\phi_{(0)}$ and $\phi_{(\Delta)}$. 

Consider now the variational principle. Varying \eqref{BulkAction} around an on-shell field configuration, one finds
\begin{equation}
\delta S\bare=-\int d^{d+1}x\sqrt{g}\left (\Box_{g}\Phi-m^{2}\Phi\right )\delta\Phi -\int_{\Sigma_r} d^{d}x\sqrt{h}\,\dot{\Phi}\delta\Phi=-\int_{\Sigma_r} d^{d}x\sqrt{h}\,\Pi\,\delta\Phi~,
\label{OnShellVariation}
\end{equation}
where in the second equality the bulk term drops out due to the equation of motion, and we expressed the boundary integral in terms of the canonically-conjugate momentum\footnote{Notice our convention for the momentum differs by a factor of $\sqrt{h}$ from the definition $\delta L/\delta\dot{\Phi}$ employed in \cite{Papadimitriou:2007sj,Papadimitriou:2010as,Papadimitriou:2003is,Papadimitriou:2004ap,Papadimitriou:2005ii,Papadimitriou:2016yit}. I.e., with respect to diffeomorphisms on $\Sigma_r$~, $\Pi$ in this paper is a scalar rather than a scalar density.\label{pidef}}
\begin{equation}
\Pi\equiv\frac{\partial \mathcal{L}}{\partial\dot{\Phi}}=\dot{\Phi}~,
\label{CanonicalMomentum}
\end{equation}
with the dot denoting the derivative with respect to $r$. The focus on $\Pi$ effects the passage to the Hamiltonian description employed in \cite{Papadimitriou:2003is,Papadimitriou:2004ap,Kraus:1999di,Papadimitriou:2005ii}, where the radial direction plays the role of time. 

The important observation is that, as one approaches $r\to\infty$, the asymptotic scaling of $h$, $\Pi$ and $\delta\Phi$ makes the right-hand side of (\ref{OnShellVariation}) $r$-dependent, implying that the variational principle \emph{at the conformal boundary} is ill defined.  
To remedy this, we ought to redefine our dynamical system by adding an appropriate boundary term $S\ct$ to the bulk action, such that $\delta(S\bare+S\ct)$ is well defined. As explained in \cite{Papadimitriou:2007sj,Papadimitriou:2010as,Papadimitriou:2016yit}, this is the fundamental origin of the need for holographic renormalization, and the fact that this procedure leads to a finite total action is a secondary, and certainly most welcome, byproduct.   

One must examine then the radial dependence of the variation of the on-shell action, and correspondingly, of the action itself. This follows from the manifest $r$-dependence in the FG expansions (\ref{FGPhi}) and (\ref{FGh}). Conventional holographic renormalization \cite{Skenderis:1999nb,deHaro:2000vlm,Skenderis:2002wp,Henningson:1998gx,Balasubramanian:1999re} first works out the divergences in terms of the FG coefficients, and then inverts the expansions to express $S\ct$ in terms of the induced fields $\Phi,h$ and their intrinsic derivatives on the cutoff surface $\Sigma_r$. A recent application of this procedure to the case of Horndeski gravity can be found in~\cite{Caceres:2023gfa}. The alternative, Hamiltonian method of \cite{Papadimitriou:2007sj,Papadimitriou:2010as,Papadimitriou:2003is,Papadimitriou:2004ap,Kraus:1999di,Papadimitriou:2005ii,Papadimitriou:2016yit} achieves the same goal through an elegant sideways move that leads directly to $S\ct[\Phi,h]$ without any need for inversion. 

To describe this, note first that the on-shell action depends only on the induced fields at the cutoff surface $\Sigma_r$.\footnote{In the special cases afflicted by the Weyl anomaly,  $S\bare\os$ also depends explicitly on $r$. This can be taken into account in the procedure without getting in the way of the main conclusions that we are choosing to summarize here \cite{Papadimitriou:2004ap,Papadimitriou:2005ii,Papadimitriou:2016yit}.}
Using the chain rule, the radial derivative of $S\bare\os\left[\Phi,h\right]$ or of any other such functional can be written as
\begin{equation}
    \partial_{r} =\int_{\Sigma_r} d^{d}x\left[ 
    \dot{h}_{ij}\frac{\delta}{\delta h_{ij}}+\dot{\Phi }\frac{\delta}{\delta \Phi } 
    \right]~.
    \label{partialr}
\end{equation}
{}From (\ref{FGh}) and (\ref{FGPhi}), we know that the leading asymptotic behavior of the induced metric and the scalar field is
\begin{subequations}
\begin{equation}
h_{ij }\sim e^{2r}h_{\left ( 0 \right )ij }\left ( x \right )~,
\label{metriasinto}
\end{equation}
\begin{equation}
\Phi \sim e^{\left ( \Delta -d \right )r}\phi_{\left ( 0 \right ) }\left ( x \right )~,
\label{escalarasinto}
\end{equation}
\end{subequations}
so we can rewrite (\ref{partialr}) as
\begin{equation}
    \partial_{r} \sim \int_{\Sigma_r} d^{d}x\left [ 2h_{\mu \nu }\frac{\delta }{\delta h_{\mu \nu }}+\left ( \Delta -d \right )\Phi \frac{\delta }{\delta \Phi } \right ]\equiv \delta _{D}~.
    \label{partialrydeltaD}
\end{equation}
The last equality defines the dilatation operator $\delta _{D}$, which acts on arbitrary functionals of the fields $\Phi$ and $h$ evaluated at the cutoff surface $\Sigma_{r}$. 
The important finding is that, to leading asymptotic order, the dilatation operator is equal to the radial derivative: 
$\delta_{D}=\partial_{r}+\mathcal{O}\left(e^{-r}\right)$.
This connection is employed by \cite{Papadimitriou:2003is,Papadimitriou:2004ap,Papadimitriou:2005ii} to replace the standard radial expansion (\ref{FGPhi}) with an expansion in eigenfunctions of the dilatation operator. This can be thought of as a rearrangement of the FG terms into covariant functionals of the induced fields whose leading $r$-dependence is specified by the corresponding dilatation eigenvalue.   

This new expansion scheme was shown to be viable and efficient within the Hamilton-Jacobi (HJ) formalism, 
which supplies two important insights: 
\begin{enumerate}[(i)]
 \item $S\bare\os$ satisfies the HJ equation
\begin{equation}
    \int d^dx\left\{\sqrt{h}\left[\frac{1}{2}\left(\frac{1}{\sqrt{h}}\frac{\delta\cS}{\delta\Phi}\right)^2-\frac{1}{2}h^{ij}\partial_i\Phi\partial_j\Phi-\frac{1}{2}m^2\Phi^2\right]+2h_{ij}\frac{\delta\cS}{\delta h_{ij}}\right\}=0~.
    \label{hj}
\end{equation} 
Renormalization in the conventional sense of constructing a set of counterterms that make the action finite can thus be understood as the task of finding a solution $\cS$ of (\ref{hj}) as an expansion in eigenfunctions of $\delta_{D}$, determining only the eigenfunctions with positive eigenvalues, which are the ones that diverge as $r\to\infty$. 
\item The canonical momentum is related to the on-shell action by\footnote{Recall footnote \ref{pidef}.} 
\begin{equation}
\frac{\delta S\bare\os}{\delta \Phi(r,x)}=\sqrt{h}\,\Pi(r,x)~.
\label{pifroms}
\end{equation}
This relation is significant because the GKPW recipe \cite{Gubser:1998bc,Witten:1998qj} equates the left-hand side with the unrenormalized one-point function $\expec{\cO}$ in the CFT (and after the addition of $S\ct$, the identification between $\expec{\cO}$ and $\Pi$ will continue to hold for their renormalized counterparts). One also learns from (\ref{pifroms}) that, just like $S\bare\os$, the canonical momentum can be regarded as a functional of the induced fields on $\Sigma_r$, subject to expansion in $\delta_{D}$-eigenfunctions on account of (\ref{partialrydeltaD}). Renormalization in the more fundamental sense of turning (\ref{OnShellVariation}) into a well-defined variational principle can thus be understood as the task of determining and subtracting from $\Pi$ all such eigenfunctions with eigenvalue larger than $-\Delta$, which are the ones that lead to divergences in (\ref{OnShellVariation}). 
\end{enumerate}

Concretely, one writes out
\begin{equation}
\Pi=\Pi_{[ d-\Delta ]}+\Pi _{[d-\Delta+2]}+\cdots +\Pi_{\left [\Delta\right ]}+\cdots~, 
\label{PiExpansionD}
\end{equation}
and similarly for $\cS$, where by definition
\begin{equation}
\delta_{D} \Pi _{\left [ s  \right ]}=-s\Pi _{\left [ s  \right ]}~.
\label{EigenDelta}
\end{equation}
{}From (\ref{escalarasinto}) and (\ref{CanonicalMomentum}), the leading term in (\ref{PiExpansionD}) is known to be 
\begin{equation}
\Pi_{[ d-\Delta ]}=\left(\Delta -d\right)\Phi~.
\end{equation}
Using this in (\ref{pifroms}) and integrating, one finds the leading term in the on-shell action,
\begin{equation}
\cS_{[ d-2\Delta ]}=\int d^dx\sqrt{h}\,\frac{\Delta -d}{2}\Phi^2~.
\end{equation}
Upon plugging the dilatation expansions of $\Pi$ and $\cS$ into (\ref{pifroms}) and (\ref{hj}) and collecting terms with the same weight, one is led to a recursive algorithm \cite{Papadimitriou:2007sj,Papadimitriou:2010as,Papadimitriou:2003is,Papadimitriou:2004ap,Kraus:1999di,Papadimitriou:2005ii,Papadimitriou:2016yit} to determine as local functions of $\Phi$ all of the $\Pi_{\left[s\right]}$ with  $s<\Delta$, and, correspondingly, all of the $\cS_{[s]}$ with $s<0$. 

One can then define
\begin{equation}
\delta S\intr\equiv-\int d^{d}x\sqrt{h}\sum_{s<\Delta }^{}\Pi _{[s]}\,\delta \Phi
\label{DilatationCounters}
\end{equation}
to ensure that the variation of the renormalized action
\begin{equation}
\delta S\ren\equiv \delta S\bare+\delta S\intr= \int d^{d}x\sqrt{h}\,\Pi _{\left [\Delta\right ]}\,\delta \Phi
\end{equation}
is finite \cite{Papadimitriou:2007sj}. The $^{\mbox{\tiny int}}$ superindex emphasizes the intrinsic nature of these counterterms, which makes transparent the connection between the variation (\ref{DilatationCounters}) and the counterterm action itself,
\begin{equation}
S\intr\equiv-\sum_{s<0 }\cS_{[s]}~.
\label{DilatationCountersAction}
\end{equation}
Explicitly, for the generic case $\Delta\neq \frac{d}{2}+n$, this reads \cite{deHaro:2000vlm,Skenderis:2002wp,Papadimitriou:2003is}
\begin{equation}
\begin{aligned}
S\intr=&\int d^dx\sqrt{h}\left(\frac{d-\Delta}{2}\Phi^2
+\frac{1}{2(2\Delta-d-2)}\Phi\Box_h\Phi\right.\\
&\qquad\qquad\quad \left.
+\frac{1}{2(2\Delta-d-2)^2(2\Delta-d-4)}\Phi\Box_h^2\Phi
+\ldots\right)~.
\end{aligned}
\label{skenderiscounterterms}
\end{equation}
This result matches the counterterms derived by the standard holographic renormalization techniques where the asymptotic expansion of the bulk action is determined by the FG expansion, and in a later stage, the divergent contributions are covariantized inverting the FG series. It becomes clear then that the introduction of the radial Hamilton-Jacobi formulation provides an efficient and elegant prescription to determine the counterterms avoiding the tedious calculations of the previous method.

In the following sections, we will explore the possibility of expressing the canonical momentum expansion in terms of eigenfunctions of the radial operator instead of the dilatation operator ones. These two sets of eigenfunctions match only to the leading order, so the effect of the subdominant terms have to be taken into account.

\section{Extrinsic Counterterms for a Massless Scalar}\label{masslesssec}

As reviewed in the previous section, the Hamiltonian method of \cite{Papadimitriou:2007sj,Papadimitriou:2010as,Papadimitriou:2003is,Papadimitriou:2004ap,Kraus:1999di,Papadimitriou:2005ii,Papadimitriou:2016yit} makes central use of the canonical momentum $\Pi=\dot{\Phi}$. \emph{A priori}, the radial derivative is  of course a quantity extrinsic to the radial cutoff surface $\Sigma_r$. Nevertheless, the method employs the on-shell relation (\ref{pifroms}) to reinterpret $\Pi$ as an \emph{intrinsic} local  functional of the fields on $\Sigma_r$. It is this reinterpretation that allows the deduction of the iterative algorithm that ultimately leads to the intrinsic counterterms (\ref{skenderiscounterterms}). The latter of course agree with the result originally obtained via FG expansion in \cite{deHaro:2000vlm}. Both methods naturally valued the reliance on purely intrinsic derivatives as a means to preserve the Dirichlet variational principle required by the GKPW recipe \cite{Gubser:1998bc,Witten:1998qj}. 

In Section \ref{Asymptotic} we have learned that, due to the conformal nature of the AdS boundary, the path integral and, consequently, the variational principle, ought to be formulated with a variation $\delta\Phi$ that is \emph{asymptotically on shell}. This in turn implies that, unlike in  systems with a standard boundary, a boundary action built with extrinsic derivatives does not in fact ruin the required Dirichlet boundary condition. Furthermore, the Hamiltonian analysis reviewed in Section \ref{Intrinsic} has brought to the forefront the relation between the conformal symmetry of the CFT at the boundary, which by definition implies invariance under dilatations, and the use of the eigenbasis of the dilatation operator to define the counterterm action by finding its kernel. It is therefore natural to consider an expansion in the eigenbasis of the radial derivative, on the grounds that, for alAdS spacetimes, the radial diffeomorphisms are directly related to Weyl rescalings at the boundary through the PBH transformation \cite{Imbimbo:1999bj,Penrose:1985bww,Brown:1986nw}.\footnote{The PBH transformation in turn preserves the FG expansion of the alAdS metric, showing that the relation between radial shifts and boundary rescalings is a kinematic property of the AdS asymptotics, with no required input from the dynamics of any particular gravity theory.} Therefore, motivated by the primacy of $\Pi$ in the elegant prescription of \cite{Papadimitriou:2007sj,Papadimitriou:2010as,Papadimitriou:2003is,Papadimitriou:2004ap,Kraus:1999di,Papadimitriou:2005ii,Papadimitriou:2016yit}, being itself a radial derivative, it is natural to wonder whether one can find an algorithm to eliminate the divergences of the bulk action using \emph{extrinsic} counterterms, which are also allowed by the variational principle. In this and the following sections, we answer this question in the affirmative.\footnote{As explained in Section~\ref{Asymptotic}, in the main text we choose to focus on counterterms that depend solely on extrinsic derivatives, leaving the `hybrid' intrinsic-extrinsic possibility for Appendix~\ref{translationapp}.}

As explained in the Introduction, we are particularly interested in the case where the scalar field is, just like the metric, massless. It is in that setting that one can expect the closest parallel between the extrinsic counterterms pursued in the present work and the Kounterterms developed to renormalize the gravitational action \cite{Olea:2005gb,Olea:2006vd,Aros:1999kt,Mora:2004rx,Miskovic:2006tm}. For this reason, in the present section we will concentrate on the $m=0$ case. Through (\ref{delta}), masslessness of the bulk scalar field implies that the dual CFT operator is marginal, $\Delta=d$.

\subsection{Massless scalar in even bulk spacetime dimension} \label{masslessgenericsubsec}

We will consider first the  case where the inequality (\ref{window}) is not saturated, which is simpler because no logarithmic terms need to be invoked in the FG expansion. Since $\Delta-\frac{d}{2}$ is required not to be an integer, the masslessness condition 
$\Delta=d$ implies that $d$ must be an odd integer, and
$n= \lfloor d/2\rfloor$.

Let us first use this paragraph to set our notation for the remainder of the paper, without assuming masslessness yet.
As we know, holographic renormalization, understood either as a 
method to obtain a well-defined variational principle \cite{Papadimitriou:2007sj} or a well-defined on-shell action \cite{deHaro:2000vlm}, requires elucidating the radial dependence of the various relevant quantities. This becomes all the more evident if we change our focus from intrinsic to extrinsic derivatives. For this reason, in what follows, instead of an expansion in terms of eigenfunctions of the dilatation operator $\delta_{D}$, we choose to work directly with the eigenfunctions of  $\partial_{r}$. The object of primary interest is the canonical momentum $\Pi$, whose radial eigenfunctions we denote as $\Pi_{\{s\}}$:
\begin{equation}
\Pi=\Pi _{\left \{ d-\Delta  \right \}}+\Pi _{\left \{ d-\Delta +2  \right \}}+\cdots +\Pi _{\left \{\Delta\right \}}+\cdots,
\label{PiExpansionr}
\end{equation}
where by definition
\begin{equation}
\partial_{r}\Pi_{\{s\}}=-s\Pi_{\{s\}}~.
\label{Eigenr}
\end{equation}
The braces in the subindices purposefully distinguish this expansion from the one labeled with brackets in (\ref{PiExpansionD}), i.e.,
$\Pi_{\{s\}}\neq\Pi_{[s]}$.  
Quantities labeled with braces or with brackets differ as well from their counterparts with parentheses, as in the Feffer-Graham expansion (\ref{FGPhi}). 
But at the same time, it is important to bear in mind that (\ref{PiExpansionr}) is just another way of writing \eqref{FGPhi}. Indeed, knowing that the general solution of the eigenfunction equation \eqref{Eigenr} is $\Pi_{\{s\}}\propto e^{-s\,r}$,  we can easily employ the radial derivative of \eqref{FGPhi} to relate  $\Pi_{\{s\}}$ to the corresponding FG coefficients $\phi_{(k)}$:
\begin{equation}
\Pi_{\{d-\Delta+k\}}=-e^{-(d-\Delta+k)r}(d-\Delta+k)\phi_{(k)}~, \quad \Pi_{\{\Delta\}}=-e^{-\Delta r}\Delta\,\phi_{(\Delta)}~.
\label{PiExact}
\end{equation}

 Now we specialize to the massless scalar in even bulk spacetime dimension, setting $\Delta=d\in 2\mathbf{N} + 1$ and $n= \lfloor d/2\rfloor$. Inspection of the FG expansion (\ref{FGPhi}) then shows that the leading term in the conjugate momentum (\ref{PiExpansionr}) vanishes, 
\begin{equation}
\Pi_{\{d-\Delta\}}=\Pi_{\{0\}}=0~.
\label{leadingpivanishes}
\end{equation}
To make progress towards our goal, we need to express $\Pi_{\{s\}}$ in terms of $\Phi$ and its radial derivatives. This can be easily done by noting that, to the relevant order for this computation, successive differentiation of (\ref{PiExpansionr}) yields
\begin{align}
\dot{\Phi}&=\Pi_{\{2\}}+\Pi_{\{4\}}+\cdots+\Pi_{\{2n\}} +\Pi _{\left \{d\right \}}+\mathcal{O}\left(e^{-(d+1)r}\right)~,
\nonumber\\
\ddot{\Phi}&=-2\Pi_{\{2\}}-4\Pi_{\{4\}}+\cdots-2n\Pi_{\{2n\}} -d\,\Pi _{\left \{d\right \}}+\mathcal{O}\left(e^{-(d+1)r}\right)~,
\nonumber\\
&\vdots
\label{PiAlgebraicSystemMassless}\\
\Phi^{(n+1)}&=(-2)^{n}\Pi _{\{2\}}+(-4)^{n}\Pi _{\{4\}}+\cdots+(-2n)^{n}\Pi _{\{2n\}}+(-d)^{n}\Pi _{\left \{d\right \}}+\mathcal{O}\left(e^{-(d+1)r}\right)~.
\nonumber
\end{align}
These are $n+1$ linear equations, from which we can algebraically solve for the $n+1$ unknowns $\Pi_{\{k\}}$ ($2\leq k\leq d$), in terms of the extrinsic derivatives of $\Phi$.\footnote{It is important to emphasize that this last statement is only true for odd $\Delta=d$, as logarithmic divergences appear for even $\Delta=d$, which need to be analyzed separately. We will discuss that case in the next subsection.}  

The novelty of the method delineated above is that, instead of inverting the FG expansion \eqref{FGPhi} and using the equation of motion \eqref{EOM} or the HJ equation (\ref{hj}) to write the coefficients $\phi_{(k)}$ as local functions of $\Phi$ and its intrinsic derivatives $\left(\Box_{h}\right)^{k}\Phi$, we solve the algebraic system of equations (\ref{PiAlgebraicSystemMassless}) to write $\Pi_{\{s\}}$ in terms of the field's extrinsic derivatives $\Phi^{(k)}$. The motivation for this is that upon invoking $\Pi$ we are already working with an extrinsic derivative, and, for the reason explained in Section \ref{Asymptotic}, there is in fact no need to translate such quantities into intrinsic ones. Unlike intrinsic derivatives, which must be contracted by pairs with the inverse induced metric for diffeomorphism invariance along the boundary, extrinsic derivatives are directly invariant.

Just like in the dilatation eigenfunction expansion reviewed in the previous section, the divergent behavior of the variation of the action originates from the terms $\Pi_{\{k\}}$ with $k<\Delta$. Consequently, aiming for a scheme with minimal subtraction, we postulate the variation of the extrinsic counterterm action to be
\begin{equation}
\delta S\ext\equiv-\int d^{d}x\sqrt{h}\left(\sum_{k=2}^{d-1}\Pi _{\left\{k \right\}}\right)\delta \Phi~.
\label{MasslessKountertermAction}
\end{equation}
The upper limit of the sum reflects the fact that $2n=d-1$. When added to (\ref{OnShellVariation}), this yields by construction the finite renormalized variation
\begin{equation}
\delta S\ren\equiv \delta S\bare+\delta S\ext= \int d^{d}x\sqrt{h}\,\Pi _{\left \{d\right \}}\,\delta \Phi
\end{equation}
On account of (\ref{metriasinto}), (\ref{escalarasinto}) and (\ref{PiExact}), this is
\begin{equation}
\delta S\ren= -d \int d^{d}x\,\phi_{(d)}\,\delta\phi_{(0)}~.
\label{MasslessRenormalizedActionEvaluated}
\end{equation}

To establish that this new renormalization method is  valid,
it remains to be shown that (\ref{MasslessKountertermAction}) is indeed the variation of a well-defined, covariant  counterterm action $S\ext$.\footnote{In accordance with the general argument of Section~\ref{introsec}, covariance of $S\ext$ under symmetries is essential to guarantee equivalence with the intrinsic approach.} Since the scalar field is involved quadratically, the natural ansatz is that (\ref{MasslessKountertermAction}) can be integrated simply by promoting $\delta\Phi\to\Phi$ and multiplying by a factor of $1/2$:  
\begin{equation}
S\ext\equiv-\frac{1}{2}\int d^{d}x\sqrt{h}\left(\sum_{k=2}^{d-1}\Pi _{\left\{k \right\}}\right) \Phi~.
\label{MasslessKountertermActionInt}
\end{equation}
And indeed, one can verify  (e.g., using the FG expansion) that the variation of (\ref{MasslessKountertermActionInt}) coincides with 
(\ref{MasslessKountertermAction}).
The covariance of our final extrinsic counterterm action (\ref{MasslessKountertermActionInt}) under boundary diffeomorphisms is consistent with the fact that the leftover numerical coefficient in the renormalized variation
(\ref{MasslessRenormalizedActionEvaluated}) is precisely the one needed for the validity of the corresponding Ward identity (see Appendix~\ref{wardapp}). 

We now list the explicit results for each odd value of $d$: 
\begin{itemize}
\item For $d=3$, $\Delta=3$ ($n=1$), one finds
\begin{equation}
\begin{aligned}
\Pi_{\{2\}}&=\frac{1}{d-2}\left(d\dot{\Phi}+\ddot{\Phi}\right),\\
\Pi_{\{3\}}&=-\frac{1}{d-2}\left(d\dot{\Phi}+\ddot{\Phi}\right),
\end{aligned}
\end{equation}
from which
\begin{equation}
\begin{aligned}
\delta S\ext&=-\int d^{d}x\sqrt{h}\,\Pi_{\{2\}}\delta\Phi\\
&=-\frac{1}{d-2}\int d^{d}x\sqrt{h}\left(d\,\dot{\Phi}+\ddot{\Phi}\right)\delta\Phi~,
\end{aligned}
\end{equation}
and consequently
\begin{equation}
S\ext=-\frac{1}{2\left(d-2\right)}\int d^{d}x\sqrt{h}\left(d\,\dot{\Phi}+\ddot{\Phi}\right)\Phi~.
\label{Sctext1ml}
\end{equation} 
\item For $d=5$, $\Delta=5$ ($n=2$), we have
\begin{equation}
\begin{aligned}
\Pi_{\{2\}}&=\frac{1}{2(d-2)}\left(4d\,\dot{\Phi}+(d+4)\ddot{\Phi}+\dddot{\Phi}\right),\\
\Pi_{\{4\}}&=-\frac{1}{2(d-4)}\left(2d\,\dot{\Phi}+(d+2)\ddot{\Phi}+\dddot{\Phi}\right),\\
\Pi_{\{5\}}&=\frac{1}{(d-4)(d-2)}\left(8\dot{\Phi}+6\ddot{\Phi}+\dddot{\Phi}\right),
\end{aligned}
\end{equation}
from which
\begin{equation}
\begin{aligned}
\delta S\ext&=-\int d^{d}x\sqrt{h}\left(\Pi_{\{2\}}+\Pi_{\{4\}}\right)\delta\Phi\\
&=-\frac{1}{\left(d-2\right)\left(d-4\right)}\int d^{d}x\sqrt{h}\left(\left(d-6\right)\dot{\Phi}-6\ddot{\Phi}-\dddot{\Phi}\right)\delta\Phi~,
\label{variSctext2}
\end{aligned}
\end{equation}
and thus
\begin{equation}
\begin{aligned}
S\ext&=-\frac{1}{2\left(d-2\right)\left(d-4\right)}\int d^{d}x\sqrt{h}\left(d\left(d-6\right)\dot{\Phi}-6\ddot{\Phi}-\dddot{\Phi}\right)\Phi~.
\label{Sctext2ml}
\end{aligned}
\end{equation}
\item For $d=7$, $\Delta=7$ ($n=3$), the calculation yields
\begin{equation}
\begin{aligned}
\Pi_{\{2\}}&=\frac{1}{8(d-2)}\left(24d\,\dot{\Phi}+(10d+24)\ddot{\Phi}+(10+d)\dddot{\Phi}+\ddddot{\Phi}\right),\\
\Pi_{\{4\}}&=-\frac{1}{4(d-4)}\left(12d\,\dot{\Phi}+(8d+12)\ddot{\Phi}+(8+d)\dddot{\Phi}+\ddddot{\Phi}\right),\\
\Pi_{\{6\}}&=\frac{1}{8(d-6)}\left(8d\dot{\Phi}+(6d+8)\ddot{\Phi}+(6+d)\dddot{\Phi}+\ddddot{\Phi}\right),\\
\Pi_{\{7\}}&=-\frac{1}{(d-6)(d-4)(d-2)}\left(48\dot{\Phi}+44\ddot{\Phi}+12\dddot{\Phi}+\ddddot{\Phi}\right),
\end{aligned}
\end{equation}
implying that
\begin{equation}
\begin{aligned}
\delta S\ext&=-\int d^{d}x\sqrt{h}\left(\Pi_{\{2\}}+\Pi_{\{4\}}+\Pi_{\{6\}}\right)\delta\Phi\\
&=-\frac{1}{\left(d-2\right)\left(d-4\right)\left(d-6\right)}\int d^{d}x\sqrt{h}\left(d\left(d\left(d-12\right)+44\right)\dot{\Phi}+44\ddot{\Phi}\right.\\
&\qquad\qquad\qquad\qquad\qquad\qquad\qquad\qquad\left.+12\dddot{\Phi}+\ddddot{\Phi}\right)\delta\Phi~,
\end{aligned}
\end{equation}
and therefore
\begin{equation}
\begin{aligned}
S\ext&=-\frac{1}{2\left(d-2\right)\left(d-4\right)\left(d-6\right)}\int d^{d}x\sqrt{h}\left(d\left(d\left(d-12\right)+44\right)\dot{\Phi}+44\ddot{\Phi}\right.\\
&\qquad\qquad\qquad\qquad\qquad\qquad\qquad\qquad\quad\left.+12\dddot{\Phi}+\ddddot{\Phi}\right)\Phi~.
\end{aligned}
\end{equation}
\end{itemize}
Thus, unlike the intrinsic counterterms series of (\ref{skenderiscounterterms}), demanding the expansion of the canonical momentum to be written in terms of eigenfunctions of the radial operator gives rise to a series of extrinsic counterterms. These are surface terms where radial derivatives of the fields become manifest and tangential derivatives are absent. In the next subsection, we generalize this prescription to the case of odd bulk dimensions.

\subsection{Massless scalar in odd bulk spacetime dimension} \label{masslessoddsubsec}

{}From (\ref{FGPhi}), we know that  an additional, linear-in-$r$ (logarithmic-in-$z$) term appears when we saturate the left inequality in (\ref{window}), i.e., when $\Delta=\frac{d}{2}+n$. In combination with the masslessness condition $\Delta=d$, we see that this happens when the CFT spacetime dimension $d$ is even,
which is also where the metric expansion in (\ref{FGh}) acquires a linear (logarithmic) term. In such case, $n=d/2$.

The modified form of the FG expansion,
\begin{equation}
\Phi=\phi _{\left ( 0 \right )}+e^{-2r}\phi _{\left (2 \right )}+\cdots +e^{-\left (d-2\right)r}\phi _{\left (d-2\right )}+e^{-d\,r}\left(\phi_{\left(d\right)}+r\tilde{\phi}_{(d)}\right)+\cdots~,
\label{FGPhiLog}
\end{equation}
implies that the conjugate momentum can no longer  be expanded purely as a linear combination of eigenfunctions of the radial derivative operator. Denoting the new term as $\tilde{\Pi}_{\left\{d\right\}}$, we have 
\begin{equation}
    \Pi =\Pi_{\left\{0 \right\}}+\Pi_{\left\{2  \right\}}+\cdots +\Pi_{\left\{d\right\}}+\tilde{\Pi}_{\left\{d  \right\}} +\cdots ~,
\label{PiExpansionrLog}
\end{equation} 
where
\begin{equation}
    \partial_{r}\Pi_{\{k\}}=-k\Pi_{\{k\}}, \qquad \partial_{r}\tilde{\Pi}_{\{d\}}\neq-d\tilde{\Pi}_{\{d\}}~.
\end{equation}
Just as in the previous subsection, we can use the radial derivative of (\ref{FGPhiLog}) to relate $\Pi_{\{s\}}$ and $\tilde{\Pi}_{\{d\}}$ to the corresponding FG coefficients $\phi_{(k)}$, which yields
\begin{equation}
\begin{aligned}
& \Pi_{\{k\}}=-e^{-kr}k\phi_{(k)}~, \\
& \Pi_{\{d\}}=e^{-d r}\left(-d\,\phi_{(d)}+\tilde{\phi}_{(d)}\right), \\
& \tilde{\Pi}_{\{d\}}=-d\, e^{-d r} r \tilde{\phi}_{(d)}~.
\end{aligned}
\label{PiExactLogMassless}
\end{equation}
Comparing to the result for the generic case (\ref{PiExact}), we can see that now the eigenfunction $\Pi_{\{d\}}$ depends on the $\tilde{\phi}_{(d)}$ coefficient, and that $\tilde{\Pi}_{\{d\}}$ fails to be an eigenfuncion of $\partial_{r}$ because of the factor linear in $r$. In fact, it is not hard to show inductively that
\begin{equation}
    (\partial_{r})^{\frac{d}{2}}\tilde{\Pi}_{\{d\}}=\left(-d\right)^{\frac{d}{2}}\left(1-\frac{1}{2 r}\right)\tilde{\Pi}_{\{d\}}~.
\end{equation}

The next step in our procedure is to express each of the $\Pi_{\{k\}}$ with $k\le d$ and $\tilde{\Pi}_{\{d\}}$ in terms of the extrinsic derivatives of the field $\Phi$. In order to do this, we can differentiate (\ref{FGPhiLog}) successively with respect to $r$, to obtain the following system of $n+1$ linear equations:
\begin{equation}
\begin{aligned}
\dot{\Phi}&=\Pi _{\{ 2\}}+\Pi _{\{ 4 \}}+\cdots +\Pi _{ \{d \}}+\tilde{\Pi} _{ \{d \}}+\mathcal{O}\left(e^{-\left ( d+2 \right )r}\right),\\
\ddot{\Phi}&=-2\Pi _{\left \{ 2  \right \}}-4\Pi _{\{ 4 \}}-\cdots -d\,\Pi _{\left \{d\right \}}+\left(-d+\frac{1}{r}\right)\tilde{\Pi} _{\left \{d\right \}}+\mathcal{O}\left(e^{-\left ( d+2 \right )r}\right),\\
&\vdots\\
\Phi^{(n+1)}&=(-2)^{n}\Pi _{\left \{0 \right \}}+(-2)^{n}\Pi _{\left \{2\right \}}+\cdots +(d)^{n}\Pi _{\left \{d  \right \}}
+(-d)^{n}\left (1-\frac{1}{2 r} \right )\tilde{\Pi} _{\left \{d\right \}}\\
&\quad +\mathcal{O}\left(e^{-\left ( d+2 \right )r}\right).
\label{PiAlgebraicSystemlogMassless}
\end{aligned}
\end{equation}
Similarly to (\ref{PiAlgebraicSystemMassless}), this algebraic system allows us to write the $n+1$ unknowns 
$\Pi_{\left\{k\right\}}$ ($k\le d$) and $\tilde{\Pi}_{\left\{d\right\}}$ in terms of the derivatives $\Phi^{\left(k\right)}$.

In parallel with the previous subsection, the variation of the extrinsic counterterm action is then postulated to be
\begin{equation}
\delta S\ext\equiv-\int d^{d}x\sqrt{h}\left(\sum_{k=2}^{d-2}\Pi _{\left\{k \right\}}+\tilde{\Pi}_{\{d\}}\right)\delta \Phi~,
\label{MasslessKountertermActionLog}
\end{equation}
so that the renormalized variation
\begin{equation}
\delta S\ren\equiv \delta S\bare+\delta S\ext= \int d^{d}x\sqrt{h}\,\Pi _{\left \{d\right \}}\,\delta \Phi
\end{equation}
is finite. 
With the aid of (\ref{metriasinto}), (\ref{FGPhiLog}) and (\ref{PiExactLogMassless}), this is
\begin{equation}
\delta S\ren=\int d^{d}x\left(-d\,\phi_{(d)}+\tilde{\phi}_{(d)}\right)\delta\phi_{(0)}~,
\label{MasslessRenormalizedActionLogEvaluated}
\end{equation}
which agrees with the result obtained via intrinsic renormalization \cite{Skenderis:2002wp}. 

Integration of  (\ref{MasslessKountertermActionLog}) again amounts simply to replacing $\delta\Phi\to\Phi$ and multiplying by a factor of $1/2$:  
\begin{equation}
S\ext\equiv-\frac{1}{2}\int d^{d}x\sqrt{h}\left(\sum_{k=2}^{d-2}\Pi _{\left\{k \right\}}+\tilde{\Pi}_{\{d\}}\right) \Phi~.
\label{MasslessKountertermActionLogInt}
\end{equation}

By construction, expression (\ref{MasslessKountertermActionLogInt}) implements minimal subtraction of the divergences in $S\bare\,$, and in particular, the term involving  $\tilde{\Pi}_{\{d\}}$ embodies a purely linear-in-$r$ (i.e., logarithimic-in-$z$) divergence. This entails that 
$\int d^{d}x\sqrt{h}(\tilde{\Pi}_{\{d\}}/r)\Phi$ is finite, and could consequently be incorporated in $S\ext$ with an arbitrary numerical coefficient. This scheme dependence is familiar from intrinsic renormalization \cite{Skenderis:2002wp}. 

Remaining within minimal subtraction, we now display the results of the extrinsic counterterm analysis for the relevant values of $n$.
 \begin{itemize}
     \item For $d=2$, $\Delta=2$ ($n=1$), solving the system (\ref{PiAlgebraicSystemlogMassless}) we find
\begin{equation}
    \begin{aligned}
 \Pi _{\left\{ 2 \right\}}&= \dot{\Phi }-r\left ( d\dot{\Phi }+\ddot{\Phi } \right ),\\
\tilde{\Pi} _{\left\{ 2 \right\}}&= r\left ( d\,\dot{\Phi }+\ddot{\Phi } \right )~,
\end{aligned}
\end{equation}
from which it follows that
\begin{equation}
\begin{aligned}
\delta S\ext&=-\int d^{d}x\sqrt{h}\,\tilde{\Pi}_{\{2\}}\delta\Phi\\
&=-\int d^{d}x\sqrt{h}\, r\left ( d\,\dot{\Phi }+\ddot{\Phi } \right )\delta\Phi~,
\end{aligned}
\end{equation}
and consequently
\begin{equation}
S\ext=-\frac{1}{2}\int d^{d}x\sqrt{h}\, r\left(d\,\dot{\Phi}+\ddot{\Phi}\right)\Phi~.
\label{Sctext1mlog}
\end{equation}

\item For $d=4$, $\Delta=4$ ($n=2$), the algebraic procedure yields
\begin{equation}
   \begin{aligned}
\Pi _{\left\{ 2 \right\}} &= 4\dot{\Phi} +2\ddot{\Phi}+\frac{1}{4}\dddot{\Phi},\\
 \Pi _{\left\{ 4 \right\}}&= -\left ( 3-4r \right )\dot{\Phi}-\left ( 2-3r \right )\ddot{\Phi}+\frac{1}{4}\left ( 1-2r \right )\dddot{\Phi},\\
 \tilde{\Pi} _{\left\{ 4 \right\}}&= -\frac{r}{2}\left ( 8\dot{\Phi}+6\ddot{\Phi}+\dddot{\Phi} \right )~,
\end{aligned}
\end{equation}
implying that 
\begin{equation}
\begin{split}
 \delta S\ext &= -\int d^d x\sqrt{h}\left(\Pi_{\{2\}}+\tilde{\Pi}_{\{4\}}\right)
 \\
 &=-\frac{1}{4}\int d^{d}x\sqrt{h}  \left[ 16\left( 1-r \right )\dot{\Phi}+4\left ( 2-3r \right )\ddot{\Phi}\right.\\
 &\qquad\qquad\qquad\qquad+\left.\left ( 1-2r \right )\dddot{\Phi} \vphantom{16\left( 1-r \right )\dot{\Phi}}\right]\,\delta\Phi~.
 \end{split}
 \end{equation}
and therefore
\begin{equation}
    \begin{split}
 S\ext&=-\frac{1}{8} \int d^{d}x\sqrt{h}\left[ 16\left ( 1-r \right )\dot{\Phi}+4\left ( 2-3r \right )\ddot{\Phi}
  +\left ( 1-2r \right )\dddot{\Phi} \right] \Phi~.
\end{split} 
\end{equation}

\item For $d=6$, $\Delta=6$ ($n=3$), one obtains
\begin{equation}
    \begin{aligned}
\Pi _{\left\{ 2 \right\}} &= \frac{1}{32}\left ( 144\dot{\Phi}+84\ddot{\Phi}+16\dddot{\Phi}+\ddddot{\Phi} \right ),\\
\Pi _{\left\{ 4 \right\}} &= \frac{1}{8}\left ( -72\dot{\Phi}-60\ddot{\Phi}-14\dddot{\Phi}-\ddddot{\Phi} \right ),\\
 \Pi _{\left\{ 6 \right\}}&= \frac{1}{32}\left [ \left ( 176-192r \right )\dot{\Phi}-4\left ( -39+44r \right )\ddot{\Phi}+\left ( 40-48r \right )\dddot{\Phi}+\left ( 3-4r \right )\ddddot{\Phi} \right ],\\
 \tilde{\Pi} _{\left\{ 6 \right\}}&= -\frac{r}{8}\left ( 48\dot{\Phi}+44\ddot{\Phi}+12\dddot{\Phi}+\ddddot{\Phi} \right )~,
\end{aligned}
\end{equation}
leading to
\begin{equation}
    \begin{aligned}
 \delta S\ext&= -\int d^d x\sqrt{h}\left(\Pi_{\{2\}}+\Pi_{\{4\}}+\tilde{\Pi}_{\{6\}}\right)
 \\
 &= \frac{1}{8}\int d^{d}x\sqrt{h}  \left [\vphantom{+\frac{1}{4}\Phi ^{\left ( 4 \right )}} 12\left ( 3-4r \right )\dot{\Phi}+\left ( 39-44r \right )\ddot{\Phi} \right.\\
&\quad\quad\quad\quad\quad\quad \quad\left.+2\left ( 5-6r \right )\dddot{\Phi}+\frac{1}{4}\left ( 3-4r \right )\ddddot{\Phi} \right]\delta\Phi
\end{aligned} 
\end{equation}
and ultimately
\begin{equation}
    \begin{aligned}
 S\ext&=\frac{1}{16}\int d^{d}x\sqrt{h}\left [ \vphantom{+\frac{1}{4}\Phi ^{\left ( 4 \right )}} 12\left ( 3-4r \right )\dot{\Phi}+\left ( 39-44r \right )\ddot{\Phi} \right.\\
&\quad\quad\quad\quad\quad\quad\quad \quad\left.+2\left ( 5-6r \right )\dddot{\Phi}+\frac{1}{4}\left ( 3-4r \right )\ddddot{\Phi} \right ] \Phi~.
\end{aligned} 
\end{equation}
 \end{itemize}
Notice that, similarly to the even{-}bulk{-}dimensional case, we obtain a series of extrinsic counterterms that depend explicitly on the radial derivatives of the scalar field. The presence of the radial coordinate in the expressions above corresponds to terms that cancel the logarithmic divergences of the bulk action. These terms break radial diffeomorphisms and are responsible for the Weyl anomaly. 

\section{Massive Scalar: generic case}
\label{Extrinsicgeneric}

We are now interested in extending the method of extrinsic renormalization to the case where the bulk scalar field $\Phi$ is massive, or equivalently, where  the dimension of its dual CFT operator  $\cO$ is $\Delta\neq d$. 
As recalled in Section~\ref{Intrinsic}, if we wish to be able to turn on a non-infinitesimal source for $\cO$, we must stick to relevant operators, with $\Delta< d$, which implies that the window-indexing integer $n$ must be chosen within the range $n< \lfloor\,d/2\, \rfloor$. 

The present section will consider the generic case where the left inequality in (\ref{window}) is not saturated, i.e., when $\Delta-\frac{d}{2}$ is not an integer. The remaining, logarithmic case, will be explored in Section~\ref{Extrinsiclog}. In both cases, the analysis will involve, as a first step, the same  algebraic procedure as in Section~\ref{masslesssec}, but there will be one important novelty. 

\subsection{Extrinsic derivative eigenfunction expansion}

Just like in the previous section, to determine $\Pi_{\{s\}}$ in terms of $\Phi$ and its radial derivatives, we successively differentiate (\ref{PiExpansionr}), obtaining
\begin{equation}
\begin{aligned}
\Phi&=\frac{1}{\Delta -d}\Pi _{\left \{d- \Delta \right \}}+\frac{1}{\Delta -d-2}\Pi _{\left \{d+2- \Delta  \right \}}+\cdots-\frac{1}{\Delta}\Pi_{\{\Delta\}}+\mathcal{O}\left(e^{(\Delta-d-2(n+1))r}\right)~,\\
\dot{\Phi}&=\Pi _{\left \{ d-\Delta  \right \}}+\Pi _{\left \{ d+2-\Delta  \right \}}+\cdots +\Pi _{\left \{\Delta\right \}}+\mathcal{O}\left(e^{(\Delta-d-2(n+1))r}\right)~,\\
&\vdots\\
\Phi^{(n+1)}&=(\Delta-d)^{n}\Pi _{\left \{d- \Delta \right \}}+(\Delta -d-2)^{n}\Pi _{\left \{d+2- \Delta  \right \}}+\cdots+(-\Delta)^{n}\Pi _{\left \{\Delta\right \}}\\
{}&\quad+\mathcal{O}\left(e^{(\Delta-d-2(n+1))r}\right)~.
\label{PiAlgebraicSystem}
\end{aligned}
\end{equation}
These are $n+2$ linear equations from which we can algebraically solve for the $n+2$ $\Pi_{\{k\}}$ with $k\leq\Delta$ in terms of $\Phi$ and its extrinsic derivatives. It is important to reiterate that we are assuming that $\frac{d}{2}+n<\Delta<\frac{d}{2}+n+1$, as logarithmic divergences appear when $\Delta=\frac{d}{2}+n$ that will be analyzed separately, in the next section. 

Just like in the previous section, the divergent behavior of the variation of the action comes from the terms $\Pi_{\{k\}}$ with $k<\Delta$. But, now that $\Delta\neq d$, an important novelty is that the minimal subtraction approach of defining $\delta S\ext\equiv-\int d^{d}x\sqrt{h}\sum_{k<\Delta}^{}\Pi _{\left\{k \right\}}$ turns out to be unviable: such variation would indeed cancel all divergences, but it cannot be integrated to a covariant counterterm action. Equivalently, it leads to a finite result that violates Ward identities, which per the general argument in Section~\ref{introsec}, is symptomatic of inequivalence with $\delta S\intr$. 

To depart from minimal subtraction, we incorporate the most general finite contribution to the variation of the action that can be built as a linear combination of any of the extrinsic derivatives $\Phi^{(k)}$. In more detail, we define 
\begin{equation}
\mathcal{F}\left(\Phi,\Phi^{(1)},\cdots,\Phi^{(n+1)}\right)=\sum_{k=0}^{n+1}f_{k}(\Delta)\Phi^{(k)}~,
\label{GeneralFiniteTerm}
\end{equation}
where by construction the numerical coefficients are to be chosen such that $\int d^{d}x\sqrt{h}\mathcal{F}\delta \Phi$ is finite. 
This condition uniquely determines all but one of the $f_{k}$, as functions of the conformal dimension $\Delta$. This can be easily seen as follows: since the leading order of the FG expansion of all the radial derivatives is the same (as it is an expansion in terms of exponential rather than polynomial functions), the boundary integral of $\mathcal{F}$ is of the form
\begin{align}
\int d^{d}x\sqrt{h}\mathcal{F}\delta\Phi&=\int d^{d}x\,e^{\Delta r}\left(e^{(\Delta-d)r}F_{(0)}+e^{(\Delta-d-2)r}F_{(2)}+\cdots\right.
\nonumber\\
&\qquad\qquad\qquad
+\left.e^{(\Delta-d-2n)r}F_{(2n)}+\cdots\right)\delta\phi_{(0)},
\end{align}
where the $F_{(k)}$ coefficients are given in terms of the $f_{k}$ and the conformal dimension $\Delta$. Hence we have $n+1$ divergent terms, that we can set to zero by fixing $n+1$ of the $f_{k}$ coefficients. We will adopt the convention of leaving $f_{0}$ undetermined (for now).

With this extra contribution at hand, we define the variation of the extrinsic counterterms action as
\begin{equation}
\delta S\ext\equiv-\int d^{d}x\sqrt{h}\left(\sum_{k<\Delta }^{}\Pi _{\left\{k \right\}}-\mathcal{F}\left(\Phi,\Phi^{(1)},\cdots,\Phi^{(n+1)}\right)\right)\delta \Phi~.
\label{KountertermAction}
\end{equation}
The variation of the renormalized action, given by\begin{equation}
\delta S\ren\equiv \delta S\bare+\delta S\ext= \int d^{d}x\sqrt{h}\left(\Pi _{\left\{\Delta\right\}}+\, \mathcal{F}\right)\delta \Phi~,
\label{RenormalizedActionKounters}
\end{equation}
is then finite by construction. While the expression for $\Pi_{\{\Delta\}}$ in terms of $\Phi^{(n)}$ will be different depending on the specific value of the conformal dimension, because of \eqref{PiExact} we know that its boundary integral will be
\begin{equation}
\int d^{d}x\sqrt{h}\Pi_{\{\Delta\}}\delta\Phi=-\Delta\int d^{d}x\,\phi_{(\Delta)}\delta\phi_{(0)}+\mathcal{O}\left(e^{(2\Delta-d-2(n+1))r}\right),
\end{equation}
for all $\Delta$ such that $\frac{d}{2}+n<\Delta<\frac{d}{2}+n+1$. Furthermore, the finite contribution in (\ref{RenormalizedActionKounters}) can also be evaluated using the FG expansion, resulting in
\begin{equation}
\int d^{d}x\sqrt{h}\,\mathcal{F}\delta\Phi=\int d^{d}x\,F_{(\Delta)}\phi_{(\Delta)}\delta\phi_{(0)}+\mathcal{O}\left(e^{(2\Delta-d-2(n+1))r}\right),
\label{FiniteTerm}
\end{equation}
where $F_{(\Delta)}$ is a function of the conformal dimension given by\footnote{This expression summarizes the results for $n$ up to 4, and is conjectured to hold in general.} 
\begin{equation}
F_{(\Delta)}=f_{0}\frac{(d-2\Delta)(d-2\Delta+2)\cdots(d-2\Delta+2n)}{(d-\Delta)(d-\Delta+2)\cdots(d-\Delta+2n)}~.
\end{equation}
As a consequence, after taking the $r\rightarrow\infty$ limit in \eqref{RenormalizedActionKounters}, we obtain
\begin{equation}
\delta S\ren= \int d^{d}x\left(-\Delta+\,F_{(\Delta)}\right)\phi_{(\Delta)}\delta\phi_{(0)}.
\label{RenormalizedActionEvaluated}
\end{equation}

\subsection{Extrinsic counterterm action\label{Sec-integrable}}
The next step is to ensure that $\delta S\ext$ as defined in (\ref{KountertermAction}) is indeed the variation of some covariant boundary action $S\ext$. Unlike what happened in the previous section, for a massive scalar field the straightforward ansatz of promoting $\delta\Phi\to\Phi$ and multiplying by $1/2$ does not work.  
We therefore start by writing down the most general boundary action that is quadratic in the field and can be built using $\Phi^{(k)}$ with combinations of up to $n+1$ derivatives,
\begin{equation}
S\ext=\int d^{d}x\sqrt{h}\,\sum_{k+l\leq n+1}\Phi^{(k)}\alpha_{k,l}\Phi^{(l)}~.
\label{SExtGeneral}
\end{equation}
We must then determine what choice of the coefficients $\alpha_{k,l}$ ensures that the variation of \eqref{SExtGeneral} coincides with \eqref{KountertermAction} order by order in the FG expansion, excluding terms that vanish when taking the $r\rightarrow\infty$ limit. 

To obtain a consistent system, we need to use the insight gained in Section~\ref{Asymptotic} that the variations we are considering are in the space of solutions, hence
\begin{equation}
\int d^{d}x\sqrt{h}\left(\phi_{(2k)}\delta\phi_{(2k-2)}-\phi_{(2k-2)}\delta\phi_{(2k)}\right)=0~,
\label{partsvariation}
\end{equation}
for any $k$, by virtue of \eqref{EOMSolution} and integration by parts. Indeed, 
\begin{equation}
\begin{aligned}
\int d^{d}x\sqrt{h}\phi_{(2k)}\delta\phi_{(2k-2)}&=\frac{e^{2r}}{2k(2\Delta-d-2k)}\int d^{d}x\sqrt{h}\,\left(\Box_{h}\phi_{(2k-2)}\right)\delta\phi_{(2k-2)}\\
&=\frac{e^{2r}}{2k(2\Delta-d-2k)}\int d^{d}x\sqrt{h}\,\phi_{(2k-2)}\Box_{h}\left(\delta\phi_{(2k-2)}\right)\\
&=\int d^{d}x\sqrt{h}\phi_{(2k-2)}\delta\phi_{(2k)}~.
\end{aligned}
\label{partsvariation2}
\end{equation}

By enforcing integrability of (\ref{KountertermAction}) through the procedure just explained, the coefficients $f_{0}$ and $\alpha_{0,n+1}$ are found to be fully determined:  
\begin{equation}
\begin{aligned}
f_{0}=&\frac{d-\Delta}{F_{(\Delta)}}=\frac{(d-\Delta)^{2}(d-\Delta+2)\cdots(d-\Delta+2n)}{(d-2\Delta)(d-2\Delta+2)\cdots(d-2\Delta+2n)}~,\\
\alpha_{0,n+1}=&\frac{1}{2(2+d-2\Delta)\cdots(2n+d-2\Delta)}~.
\end{aligned}
\end{equation}
Additionally, the terms in (\ref{SExtGeneral}) that multiply the undetermined $\alpha_{k,l}$ are found in the end to vanish in the  $r\rightarrow\infty$ limit. This means that we have many equivalent ways of writing the extrinsic counterterm action, depending on the arbitrary coefficients that multiply these non-obviously vanishing contributions (this will become clearer in the example of the next subsection). Nevertheless, the final evaluation of the on-shell renormalized action will always be
\begin{equation}
S\ren=S\bare+S\ext=-(2\Delta-d)\int d^{d}x\,\phi_{(0)}\phi_{(\Delta)}~,
\label{finalsren}
\end{equation}
consistent with the results obtained using other holographic renormalization methods \cite{deHaro:2000vlm,Skenderis:2002wp,Papadimitriou:2010as,Papadimitriou:2003is,Papadimitriou:2004ap}. In particular, as reviewed in Appendix~\ref{wardmassiveapp}, the $2\Delta -d$ coefficient in (\ref{finalsren}), which our method fixes uniquely by demanding integrability of the variation (\ref{KountertermAction}), is precisely the one that is known to be required to satisfy Ward identities. From this vantage point, we see that what is special about the case of a massless bulk field, analyzed in the previous section, is that it is only for $\Delta=d$ that the 
renormalized variation of the action
produced by minimal subtraction, (\ref{MasslessRenormalizedActionEvaluated}),
has a numerical coefficient $d$ that directly agrees with the value required by Ward identities, $2\Delta -d$.  It is for this reason that extrinsic renormalization can be carried out with maximal ease when the bulk field is massless. 

\subsection{An example}
In order to illustrate the construction of the extrinsic counterterm action for a massive bulk scalar field, we will now show the explicit computation for a conformal dimension $\Delta$ such that $\frac{d}{2}+1<\Delta<\frac{d}{2}+2$ (that is, $n=1$ in the notation of (\ref{ndef})), following all the steps that we previously described. To the relevant order for this computation, the FG expansion of the field is given by
\begin{equation}
\Phi=e^{-\left (d-\Delta\right )r}\phi _{\left(0\right)}+e^{-\left(d-\Delta+2\right )r}\phi _{\left(2\right)}+e^{-\Delta r}\phi_{\left(\Delta\right)}+\mathcal{O} \left(e^{-\left(d-\Delta+4\right )r} \right ).
\label{FGExample}
\end{equation}

For this conformal dimension, the algebraic system of equations \eqref{PiAlgebraicSystem} takes the form
\begin{equation}
\begin{aligned}
 \Phi &= \frac{1}{\Delta -d}\Pi _{\left \{d- \Delta \right \}}+\frac{1}{\Delta -d-2}\Pi _{\left \{d+2- \Delta  \right \}}-\frac{1}{\Delta }\Pi _{\left \{ \Delta  \right \}}+\mathcal{O} \left ( e^{\left ( \Delta -d -4\right )r} \right ), \nonumber\\
 \dot{\Phi }&= \Pi _{\left \{d- \Delta \right \}}+\Pi _{\left \{ d+2-\Delta  \right \}}+\Pi _{\left \{ \Delta  \right \}}+\mathcal{O} \left ( e^{\left ( \Delta -d -4\right )r} \right ),\\
 \ddot{\Phi}&= \left ( \Delta-d \right )\Pi _{\left \{ d-\Delta \right \}}+\left ( \Delta -d-2 \right )\Pi _{\left \{d+2- \Delta \right \}}-\Delta \Pi _{\left \{ \Delta  \right \}}+\mathcal{O} \left ( e^{\left ( \Delta -d -4\right )r} \right )~.
\end{aligned}
\label{AlgebraicSystemExampleN1}
\end{equation}
{}From here we can solve for the radial derivative eigenfunctions $\Pi_{\{n\}}$ in terms of the extrinsic derivatives of the field:
\begin{equation}
\begin{aligned}
& \Pi _{\left \{d-\Delta \right \}}=\frac{\Delta(d-\Delta )(d-\Delta +2)}{2 (d-2 \Delta )}\Phi+\frac{(d+2) (d-\Delta )}{2 (d-2 \Delta )}\dot{\Phi}+\frac{d-\Delta }{2 d-4 \Delta }\ddot{\Phi}~,\\
& \Pi _{\left \{d+2- \Delta  \right \}}=-\frac{\Delta  (d-\Delta ) (d-\Delta +2)}{2 (d-2 \Delta +2)}\Phi-\frac{d (d-\Delta +2)}{2 (d-2 \Delta +2)}\dot{\Phi}-\frac{d-\Delta +2}{2 d-4 \Delta +4}\ddot{\Phi}~,\\
& \Pi _{\left \{ \Delta  \right \}}=-\frac{\Delta  (d-\Delta ) (d-\Delta +2)}{(d-2 \Delta ) (d-2 \Delta +2)}\Phi-\frac{2 \Delta  (d-\Delta +1)}{(d-2 \Delta ) (d-2 \Delta +2)}\dot{\Phi}-\frac{\Delta }{(d-2 \Delta ) (d-2 \Delta +2)}\ddot{\Phi}~.
\end{aligned}
\label{PiExampleN1}
\end{equation}

The most general additional integrand yielding a finite contribution to $\delta S\ext$ in (\ref{KountertermAction}) is of the form
\begin{equation}
\mathcal{F}\left(\Phi,\dot{\Phi},\ddot{\Phi}\right)=f_{0}\Phi+f_{1}\dot{\Phi}+f_{2}\ddot{\Phi}~.
\end{equation}
Substituting the FG expansion \eqref{FGExample}, and demanding that the boundary integral \eqref{FiniteTerm} remain finite fixes $f_{1}$ and $f_{2}$ as functions of $f_{0}$ and $\Delta$. Explicitly, we obtain
\begin{equation}
\mathcal{F}\left(\Phi,\dot{\Phi},\ddot{\Phi}\right)=f_{0}\left(\Phi+\frac{2(d-\Delta+1)}{(d-\Delta)(d-\Delta +2)}\dot{\Phi}+\frac{1}{(d-\Delta ) (d-\Delta +2)}\ddot{\Phi}\right).
\label{FiniteTermExampleN1}
\end{equation}

By substitution of \eqref{PiExampleN1} and \eqref{FiniteTermExampleN1} in \eqref{KountertermAction}, we arrive at 
\begin{equation}
\begin{aligned}
\delta S\ext=&-\int d^{d}x\sqrt{h}\left(\frac{\Delta \left ( \Delta -d \right )\left ( \Delta -d-2 \right )}{\left ( 2\Delta -d \right )\left ( 2\Delta -d-2 \right )}-f_{0}\right)\Phi\,\delta\Phi\\
&-\int d^{d}x\sqrt{h}\left(\frac{\left ( \Delta -d \right )\left ( \Delta -d-2 \right )+\Delta ^{2}}{\left ( 2\Delta -d \right )\left ( 2\Delta -d-2 \right )}+f_{0}\frac{2\left ( \Delta -d-1 \right )}{\left ( \Delta -d \right )\left ( \Delta -d-2 \right )}\right)\dot{\Phi}\,\delta\Phi\\
&-\int d^{d}x\sqrt{h}\left(\frac{\Delta }{\left ( 2\Delta -d \right )\left ( 2\Delta -d-2 \right )} -\frac{f_{0} }{\left ( \Delta -d \right )\left ( \Delta -d-2 \right )}\right)\ddot{\Phi}\,\delta\Phi~.
\end{aligned}
\label{VariationExample}
\end{equation}
As we can explicitly check using \eqref{FGPhi}, the variation of the renormalized action \eqref{RenormalizedActionKounters} is finite in the  $r\rightarrow\infty$ limit, yielding
\begin{equation}
\delta S\ren=\int d^{d}x\left(-\Delta+f_{0}\frac{(d-2 \Delta ) (d-2 \Delta +2) }{(d-\Delta ) (d-\Delta +2)}\right)\phi_{(\Delta)}\delta\phi_{(0)}~.
\end{equation}
That is, in the notation of \eqref{RenormalizedActionEvaluated}, for this range of values of the conformal dimension we have
\begin{equation}
F_{(\Delta)}=f_{0}\frac{(d-2 \Delta ) (d-2 \Delta +2) }{(d-\Delta ) (d-\Delta +2)}~.
\end{equation}

As explained previously, the final step is to demand that $\delta S\ext$ is truly the variation of a boundary action. In the present case, we expect $S\ext$ to be a quadratic action of the field with up to two radial derivatives,
\begin{equation}
S\ext=\int d^{d}x\sqrt{h}\left(\Phi\alpha_{0,0}\Phi+\Phi\alpha_{0,1}\dot{\Phi}+\Phi\alpha_{0,2}\ddot{\Phi}+\dot{\Phi}\alpha_{1,1}\dot{\Phi}\right)~,
\end{equation}
whose variation is
\begin{equation}
\begin{aligned}
\delta S\ext=&\int d^{d}x\sqrt{h}\,\left(2\Phi\delta\Phi\,\alpha_{0,0}+\left(\dot{\Phi}\delta\Phi+\Phi\delta\dot{\Phi}\right)\alpha_{0,1}+\left(\ddot{\Phi}\delta\Phi+\Phi\delta\ddot{\Phi}\right)\alpha_{0,2}+2\dot{\Phi}\delta\dot{\Phi}\,\alpha_{1,1}\right).
\end{aligned}
\label{VariationGeneralExample}
\end{equation}
Substituting the FG expansion \eqref{FGExample} in \eqref{VariationExample} and \eqref{VariationGeneralExample}, and equating the coefficients of the divergent and finite terms proportional to $\phi_{(k)}\delta\phi_{(l)}$ in both, we obtain a system of algebraic equations that we can solve for $\alpha_{k,l}$. 

More specifically, equating the coefficients of the terms proportional to $\phi_{(0)}\delta\phi_{(0)}$, $\phi_{(0)}\delta\phi_{(\Delta)}$, $\phi_{(\Delta)}\delta\phi_{(0)}$, and $\phi_{(2)}\delta\phi_{(0)}$, respectively, yields
\begin{equation}
\begin{aligned}
0&=2\alpha_{0,0}+2\Delta(\alpha_{0,1}+\Delta(\alpha_{0,2}+\alpha_{1,1}))+2d^{2}(\alpha_{0,2}+\alpha_{1,1})-d(2\alpha_{0,1}+4\Delta(\alpha_{0,2}+\alpha_{1,1}))\\
&\quad-d+\Delta~,\\
0&=2\alpha_{0,0}+2\Delta^{2}(\alpha_{0,2}-\alpha_{1,1})+\alpha_{0,2}d^{2}-d(\alpha_{0,1}+2\alpha_{0,2}\Delta-2\alpha_{1,1}\Delta +1)+\Delta~,\\
0&=2\alpha_{0,0}+2\Delta^{2}(\alpha_{0,2}-\alpha_{1,1})-f_{0}\frac{(d-2\Delta)(d-2\Delta+2)}{(d-\Delta)(d-\Delta+2)}+d(\alpha_{0,2}d-\alpha_{0,1})\\
&\quad+2d\Delta(\alpha_{1,1}-\alpha_{0,2})~,\\
0&=2(\alpha_{0,0}-\alpha_{0,1})+2\Delta^{2}(\alpha_{0,2}+\alpha_{1,1})+4\alpha_{0,2}+2\Delta(\alpha_{0,1}-2(d+1)(\alpha_{0,2}+\alpha_{1,1}))\\
&\quad+d(-2\alpha_{0,1}+2\alpha_{0,2}(d+2)+2\alpha_{1,1}(d+2)-1)+\Delta -1~.
\end{aligned}
\end{equation}
The general solution can be written as
\begin{equation}
\begin{aligned}
f_{0}=&\frac{(d-\Delta )^2 (d-\Delta +2)}{(d-2 \Delta ) (d-2 \Delta +2)},\\
\alpha_{0,2}=&\frac{1}{2 d-4 \Delta +4},\\
\alpha_{0,1}=&\frac{d}{2 d-4 \Delta +4}+2\alpha_{1,1}(d-\Delta),\\
\alpha_{0,0}=&\frac{(d-\Delta ) (d-\Delta +2)}{2 (d-2 \Delta +2)}+\alpha_{1,1}(d-\Delta )^{2}
\end{aligned}
\end{equation}
To arrive at these results we used the fact that, as explained around (\ref{partsvariation}) and (\ref{partsvariation2}), $\int d^{d}x\sqrt{h}\left(\phi_{0}\delta\phi_{2}-\phi_{2}\delta\phi_{0}\right)=0$, by virtue of the equation of motion (because the variations we are considering are asymptotically in the space of solutions). 

As expected, imposing the integrability condition fixed $f_{0}$ and the finite contribution to the extrinsic counterterms. We are then left with
\begin{equation}
\begin{aligned}
S\ext=&\int d^{d}x\,\sqrt{h}\,\Phi\left(\frac{(d-\Delta)(d-\Delta+2)}{2(d-2\Delta +2)}\Phi+\frac{d}{2d-4\Delta+4}\dot{\Phi}+\frac{1}{2d-4\Delta+4}\ddot{\Phi}\right)\\
&+\alpha_{1,1}\int d^{d}x\,\sqrt{h}\,\left(\dot{\Phi}+(d-\Delta)\Phi\right)^{2}~.
\end{aligned}
\label{sextwithalpha11}
\end{equation}
The contribution of the free coefficient $\alpha_{1,1}$ consistently vanishes when we take the  $r\rightarrow\infty$ limit. We are thus free to choose any value for it, thereby changing the appearance of $S\ext$. 

One option is to choose  $\alpha_{1,1}=0$, in which case
\begin{equation}
S\ext=\int d^{d}x\,\sqrt{h}\,\Phi\left(\frac{(d-\Delta)(d-\Delta+2)}{2(d-2\Delta +2)}\Phi+\frac{d}{2d-4\Delta+4}\dot{\Phi}+\frac{1}{2d-4\Delta+4}\ddot{\Phi}\right)~.
\label{Sctextfirstwindow}
\end{equation}
Upon replacing the extrinsic derivatives with $\Box_{h}$ via the equation of motion, this coincides with the intrinsic counterterms (\ref{skenderiscounterterms}). 

Alternative choices of $\alpha_{1,1}$ in (\ref{sextwithalpha11}) can be made to eliminate either the $\Phi^2$, or $\Phi\dot{\Phi}$, or $\Dot{\Phi}^2$ counterterms. However, independently of the value of $\alpha_{1,1}$, the final result for the on-shell renormalized action is
\begin{equation}
S\ren=S\bare+S\ext=-(2\Delta-d)\int d^{d}x\,\phi_{(0)}\phi_{(\Delta)}~,
\end{equation}
consistent with the results obtained using other holographic renormalization methods \cite{deHaro:2000vlm,Papadimitriou:2010as,Papadimitriou:2004ap}.

\subsection{Abbreviated form of the extrinsic renormalization method}\label{abreviated}
In the preceding subsections, we have presented the method of extrinsic renormalization for a massive scalar field in a form that is originally motivated by the Hamiltonian method of   \cite{Papadimitriou:2007sj,Papadimitriou:2010as,Papadimitriou:2003is,Papadimitriou:2004ap,Papadimitriou:2005ii,Papadimitriou:2016yit}, and that throws light on the special nature of the massless case analyzed in Section~\ref{masslesssec}. In this presentation, the method has required two additional steps beyond the purely algebraic computation that suffices when the field is massless. Knowing this, it is more efficient to reformulate the procedure by shifting our focus back from the variation of the action  to the action itself. We can then simply start from the general extrinsic counterterm action (\ref{SExtGeneral}) and use the FG expansion to choose the numerical coefficients in such a way as to cancel all of the divergences in $S\bare$. The overall procedure is then abbreviated from three steps down to only one. 

In this guise, extrinsic renormalization parallels the original method of \cite{deHaro:2000vlm,Skenderis:2002wp}, with the single modification of using counterterms built with extrinsic instead of intrinsic derivatives. Such abbreviated procedure makes no mention of the conjugate momentum $\Pi$, and is consequently devoid of conceptual and structural contact with the more elegant method of \cite{Papadimitriou:2007sj,Papadimitriou:2010as,Papadimitriou:2003is,Papadimitriou:2004ap,Papadimitriou:2005ii,Papadimitriou:2016yit}. From the practical standpoint, this is not much loss with respect to the previous subsections, because already there we had made use of the FG expansion, which is a feature that \cite{deHaro:2000vlm,Skenderis:2002wp} efficiently avoided by the Hamiltonian method.

\section{Massive Scalar: logarithmic case}
\label{Extrinsiclog}
\subsection{General features}
As emphasized in the previous section, the 
massive (i.e., $\Delta\neq d$) case 
with $\Delta=\frac{d}{2}+n$ must be treated separately, as some new features appear that need special care. In this case, the FG expansion of the scalar field $\Phi$ acquires an additional term at order $e^{-\Delta r}$, namely
\begin{equation}
\Phi=e^{-\left (d-\Delta\right )r}\phi _{\left ( 0 \right )}+e^{-\left (d-\Delta+2\right)r}\phi _{\left (2 \right )}+\cdots +e^{-\left (d-\Delta+2n-2\right)r}\phi _{\left (2n-2\right )}+e^{-\Delta r}\left(\phi_{\left(\Delta\right)}+r\tilde{\phi}_{(\Delta)}\right)+\cdots.
\label{FGPhiLog2}
\end{equation}
 As a direct consequence of this, the conjugate momentum can no longer be expanded as a linear combination of eigenfunctions of the radial derivative operator, and in turn an extra term, which we will denote as $\tilde{\Pi}_{\left\{\Delta\right\}}$, is necessary in order to achieve an expansion. Explicitly, we have 
\begin{equation}
    \Pi =\Pi_{\left\{d-\Delta  \right\}}+\Pi_{\left\{d-\Delta+2  \right\}}+\cdots +\Pi_{\left\{\Delta  \right\}}+\tilde{\Pi}_{\left\{\Delta  \right\}} +\cdots ~,
\label{PiExpansionrLogM}
\end{equation} 
where
\begin{equation}
    \partial_{r}\Pi_{\{s\}}=-s\Pi_{\{s\}}, \qquad \partial_{r}\tilde{\Pi}_{\{\Delta\}}\neq-s\tilde{\Pi}_{\{\Delta\}}.
\end{equation}
Just as in the generic case of the previous section, we can use the radial derivative of \eqref{FGPhi} to relate $\Pi_{\{s\}}$ and $\tilde{\Pi}_{\{\Delta\}}$ to the corresponding FG coefficients $\phi_{(k)}$, which yields
\begin{equation}
\begin{aligned}
& \Pi_{\{d-\Delta+k\}}=-e^{-(d-\Delta+k)r}(d-\Delta+k)\phi_{(k)}~, \\
& \Pi_{\{\Delta\}}=e^{-\Delta r}\left(-\Delta\,\phi_{(\Delta)}+\tilde{\phi}_{(\Delta)}\right), \\
& \tilde{\Pi}_{\{\Delta\}}=-\Delta e^{-\Delta r} r \tilde{\phi}_{(\Delta)}~.
\end{aligned}
\label{PiExactLog}
\end{equation}
Comparing to the result for the generic case, (\ref{PiExact}), we can see that now the eigenfunction $\Pi_{\{\Delta\}}$ depends on the $\tilde{\phi}_{(\Delta)}$ coefficient, and that $\tilde{\Pi}_{\{\Delta\}}$ fails to be an eigenfuncion of $\partial_{r}$ because of the factor linear in $r$ (logarithmic in $z$). Using mathematical induction over $n$, it is not hard to show   that
\begin{equation}
    (\partial_{r})^{n}\tilde{\Pi}_{\{\Delta\}}=\left(-\Delta\right)^{n}\left(1-\frac{n}{\Delta r}\right)\tilde{\Pi}_{\{\Delta\}}~.
\end{equation}

As always, the next step in our procedure is to express each of the $\Pi_{\{s\}}$ and $\Pi_{\{\Delta\}}$ in terms of the field $\Phi$ and its radial derivatives. In order to do this, we can differentiate \eqref{PiExpansionrLogM} successively with respect to $r$, to obtain the following system of $n+2$ linear equations:
\begin{equation}
\begin{aligned}
\Phi&=\frac{1}{\Delta -d}\Pi _{\left \{d- \Delta \right \}}+\frac{1}{\Delta -d+2}\Pi _{\left \{d- \Delta +2\right \}}+\cdots -\frac{1}{\Delta}\Pi _{\left \{ \Delta  \right \}}\\
&\quad -\frac{1}{\Delta}\left ( 1+\frac{1}{\Delta r} \right )\tilde{\Pi}_{\{\Delta\}}+\mathcal{O}\left(e^{-\left ( \Delta+2 \right )r}\right),\\
\dot{\Phi}&=\Pi _{\left \{ d-\Delta  \right \}}+\cdots +\Pi _{\left \{\Delta\right \}}+\tilde{\Pi} _{\left \{\Delta\right \}}+\mathcal{O}\left(e^{-\left ( \Delta+2 \right )r}\right),\\
&\vdots\\
\Phi^{(n+1)}&=(\Delta-d)^{n}\Pi _{\left \{d- \Delta \right \}}+(\Delta-d+2)^{n}\Pi _{\left \{d- \Delta +2\right \}}+\cdots +(-\Delta)^{n}\Pi _{\left \{ \Delta  \right \}}\\
&\quad +(-\Delta)^{n}\left (1-\frac{n}{\Delta r} \right )\tilde{\Pi} _{\left \{\Delta\right \}}+\mathcal{O}\left(e^{-\left ( \Delta+2 \right )r}\right).
\label{PiAlgebraicSystemlog}
\end{aligned}
\end{equation}
Similarly to (\ref{PiAlgebraicSystem}), this algebraic system allows us to write the $\Pi_{\left\{k\right\}}$ and $\tilde{\Pi}_{\{\Delta\}}$ in terms of $\Phi$ and its extrinsic derivatives $\Phi^{\left(k\right)}$.

To complete the variation of the extrinsic counterterm action (\ref{KountertermAction}), we note that the most general finite contribution $\mathcal{F}$ now reads 
\begin{equation}
\mathcal{F}\left(\Phi,\Phi^{(1)},\cdots,\Phi^{(n+1)}\right)=\sum_{k=0}^{n+1}\left(f_{k}(\Delta)+\tilde{f}_{k}(\Delta)r\right)\Phi^{(k)},
\label{GeneralFiniteTermLog}
\end{equation}
where $f_{k}$ and $\Tilde{f}_{k}$ are arbitrary coefficients that depend on $\Delta.$ As in the generic case of Section \ref{Extrinsicgeneric}, we obtain a general coefficient for the finite contribution $\mathcal{F},$ given by
\begin{equation}
     \int d^{d}x\sqrt{h}\mathcal{F}\delta \Phi =\int d^{d}x\left ( \tilde{F}_{\left ( \Delta  \right )}\phi _{\left ( \Delta  \right )}+C\left ( \phi _{\left ( 0 \right )} \right )\tilde{\phi }_{\left ( \Delta  \right )} \right )\delta \phi _{\left ( 0 \right )}+\mathcal{O}\left ( e^{-\left ( \Delta +2 \right )r} \right ),
\end{equation}
where $C\left ( \phi _{\left ( 0 \right )}\right)$ is a scheme dependent coefficient and $\tilde{F}_{\left ( \Delta  \right )}$ reads
\begin{equation}
    \tilde{F}_{\left ( \Delta  \right )}=\frac{\left ( -1 \right )^{n+1}2^{n}n!}{\left ( d-\Delta  \right )\left ( 2+d-\Delta  \right )\cdots \left ( 2n+d-\Delta  \right )}\tilde{f}_{0}.
\end{equation}

To be able to enforce integrability of the variation (\ref{KountertermAction}), the extrinsic counterterm action (\ref{SExtGeneral}) must now be generalized to have $2(n+2)$ coefficients,
\begin{equation}
S\ext=\int d^{d}x\sqrt{h}\,\sum_{k+l\leq n+1}\Phi^{(k)}\left(\alpha_{k,l}+\tilde{\alpha}_{k,l}r\right)\Phi^{(l)}~.
\end{equation}
After imposing the integrability condition explained in Sec.~\ref{Sec-integrable}, the following coefficients are fully determined:\footnote{This expression summarizes the results for $n$ up to 4, and is conjectured to hold in general.} 
\begin{equation}
    \tilde{\alpha }_{0,n+1}=\frac{\left ( -1 \right )^{n}n}{2^{n}n!},
\end{equation}
and 
\begin{equation}
    \tilde{f }_{0}=\frac{d-\Delta }{\tilde{F}_{\left ( \Delta  \right )}}.
\end{equation}

\subsection{An example}
To illustrate the procedure just described, we will compute the extrinsic counterterm action explicitly for the simplest case, $n=1$. The FG expansion reads
\begin{equation}
\Phi =e^{-\left (d-\Delta\right )r}\phi _{\left ( 0 \right )}+e^{-\Delta r}\left(\phi_{(\Delta)}+r\tilde{\phi}_{(\Delta)}\right)+\mathcal{O} \left(e^{-\left(\Delta+2\right )r} \right )~,
\label{logaFG1}
\end{equation}
and in turn the expansion of $\Pi$ is
\begin{equation}
\Pi=\Pi _{\left\{d-\Delta\right\}}+\Pi_{\{\Delta\}}+\Tilde{\Pi} _{\left\{ \Delta\right\}}+\ldots~.
\end{equation}
We emphasize that $\tilde{\Pi}$ is not an eigenfunction of the $\partial_{r}$ operator. The system of algebraic equations that needs to be solved for $\Pi_{\{k\}}$ and $\tilde{\Pi}_{\{\Delta\}}$ is
\begin{equation}
 \begin{aligned}
\Phi  &=\frac{1}{\Delta -d}\Pi _{\left\{ d-\Delta \right\}}+\frac{1}{\Delta -d-2}\Pi _{\left\{ \Delta \right\}} +\left ( 1-\frac{1}{\left ( \Delta -d-2 \right )r} \right )\tilde{\Pi} _{\left\{ \Delta \right\}}+\mathcal{O} \left(e^{-\left(\Delta+2\right )r} \right ),\\
\dot{\Phi } &= \Pi _{\left\{ d-\Delta \right\}}+\Pi _{\left\{ \Delta \right\}} +\tilde{\Pi} _{\left\{ \Delta \right\}}+\mathcal{O} \left(e^{-\left(\Delta+2\right )r} \right ),\\
\ddot{\Phi } &= \left ( \Delta -d \right )\Pi _{\left\{ d-\Delta \right\}}+\left ( \Delta -d-2 \right )\Pi _{\left\{ \Delta \right\}}+\left ( \frac{1}{r}+\Delta -d-2 \right )\tilde{\Pi} _{\left\{ \Delta \right\}}+\mathcal{O} \left(e^{-\left(\Delta+2\right )r} \right ).
\end{aligned}
\label{systemPhin1}
\end{equation}
The solution (with $\Delta=\frac{d}{2}+1$) is
\begin{equation}
     \begin{aligned}
\Pi _{\left\{ d-\Delta \right\}} &=\frac{\left ( d-2 \right )}{32}\left [ \left ( d+2 \right )^{2}\Phi  +4\left ( d+2 \right )\dot{\Phi }+4\ddot{\Phi }\right ]~, \\
\tilde{\Pi } _{\left\{ \Delta \right\}} &=\frac{\left ( d+2 \right )}{16}\left [ \left ( d^{2}-4 \right )\Phi +4\left ( d\dot{\Phi }+\ddot{\Phi } \right ) \right ] r~,\\ 
\Pi _{\left\{ \Delta \right\}} &=\frac{1}{32}\left [ \left ( d+2 \right )\left ( d^{2}-4 \right )\Phi  +4\left ( d^{2}+4 \right )\dot{\Phi }+4\left ( d-2 \right )\ddot{\Phi }\right ] \\
 &\quad -\frac{\left ( d+2 \right )}{16}\left [ \left ( d^{2}-4 \right )\Phi +4\left ( d\dot{\Phi }+\ddot{\Phi } \right ) \right ] r~.
\end{aligned}
\end{equation}

The most general finite contribution to (\ref{KountertermAction}) is of the form
\begin{equation}
\mathcal{F}\left(\Phi,\dot{\Phi},\ddot{\Phi}\right) = \left ( f_{0}\Phi +f_{1}\dot{\Phi }+f_{2}\ddot{\Phi } \right )+r\left ( \tilde{f}_{0}\Phi +\tilde{f}_{1}\dot{\Phi }+\tilde{f}_{2}\ddot{\Phi } \right )~.
\label{Flog}
\end{equation}
Employing \eqref{logaFG1} and demanding finiteness of \eqref{FiniteTerm}, four of the six coefficients in (\ref{Flog}) are fixed in terms of the remaining two. Choosing $f_{0}$ and $\tilde{f}_{0}$ as our free parameters, we obtain
\begin{equation}
\begin{aligned}
\mathcal{F}\left(\Phi,\dot{\Phi},\ddot{\Phi}\right)=&\left(f_{0}+r\tilde{f}_{0}\right)\left(\Phi+\frac{4d}{d^{2}-4}\dot{\Phi }+\frac{4}{d^{2}-4}\ddot{\Phi } \right )\\
&+\tilde{f}_{0}\left(\frac{4}{\left ( d+2 \right )^{2}}\dot{\Phi }+\frac{8}{\left ( d+2 \right )^{2}\left ( d-2 \right )}\ddot{\Phi }\right)~.
\end{aligned}
\end{equation}
The complete variation of the extrinsic counterterm action is thus
\begin{equation}
     \begin{aligned}
\delta S\ext &=\int d^{d}x\sqrt{h}\left [ \left ( \frac{\left ( d-2 \right )\left ( d+2 \right )^{2}\left ( 1-2r \right )}{32}+f_{0}+\tilde{f}_{0} \right )\Phi  \right. \\
 &\quad\quad\quad\quad\quad\quad +\left ( \frac{\left ( d+2 \right )\left ( d-2-2dr \right )}{8}+\frac{4d}{d^{2}-4}f_{0}+\frac{4\left ( d-2+d\left ( d+2 \right )r \right )}{\left ( d+2 \right )\left ( d^{2}-4 \right )}\tilde{f}_{0} \right )\dot{\Phi }\\
 &\quad\quad\quad\quad\quad\quad +\left. \left ( \frac{\left ( d-2 \right )-2\left ( d+2 \right )r}{8}+\frac{4}{d^{2}-4}f_{0}+\frac{4\left ( 2+\left ( d+2 \right )r \right )}{\left ( d+2 \right )\left ( d^{2}-4 \right )}\tilde{f}_{0} \right )\ddot{\Phi } \right ]\delta \Phi .
\end{aligned}
\label{deltas_extlog1}
\end{equation}
The procedure ensures that the variation of the renormalized action is finite, and it is given by
\begin{equation}
\delta S\ren=-\int d^{d}x\left [\frac{\left(d+2 \right)^{2}\left (d-2\right)-16\tilde{f}_{0}}{2\left(d^{2}-4\right)}\phi _{\left(\Delta\right)}+\left(2-\frac{16f_{0}}{d^{2}-4}-\frac{8\left(d+6\right)\tilde{f}_{0}}{\left(d+2\right)^{2}\left(d-2\right)}\right)\tilde{\phi }_{\left (\Delta\right)} \right ]\delta \phi _{\left ( 0 \right )}. 
\label{srengammasigma}
\end{equation}

As previously explained, $\delta S\ext$ is meant to be the variation of a boundary action, expected to be of the form
\begin{equation}
S\ext=\int d^{d}x\sqrt{h}\,\Phi\sum_{k=0}^{2}\left(\alpha_{0,k}+\Tilde{\alpha}_{0,k}r\right)\Phi^{(k)}+\int d^{d}x\sqrt{h}\,\dot{\Phi}\left(\alpha_{1,1}+\Tilde{\alpha}_{1,1}r\right)\dot{\Phi}~.
\label{logKtaction}
\end{equation}
Substituting the FG expansion \eqref{logaFG1} in \eqref{deltas_extlog1} and in the variation of \eqref{logKtaction}, we can again equate the coefficients of the divergent and finite terms proportional to $\phi_{(k)}\delta\phi_{(l)}$ in both, obtaining a system of algebraic equations for $\alpha_{k,l}$. The coefficients of the terms proportional to $\phi_{(0)}\delta\phi_{(0)}$, $\phi_{(0)}\delta\phi_{(\Delta)}$, $\phi_{(0)}\delta\tilde{\phi}_{(\Delta)}$,
$\phi_{(\Delta)}\delta\phi_{(0)}$,
$\phi_{(\Delta)}\delta\phi_{(\Delta)}$,
$\phi_{(\Delta)}\delta\tilde{\phi}_{(\Delta)}$ and $\tilde{\phi}_{(\Delta)}\delta\phi_{(0)}$ yield respectively
\begin{equation}
\begin{aligned}
d-2 &= 4\alpha_{0,0}+2\left ( 2-d \right )\alpha_{0,1}+\left ( d-2 \right )^{2}\alpha_{0,2}+\left ( 2-d \right )^{2}\alpha_{1,1}~, \\
0 &=4\Tilde{\alpha}_{0,0}+2\left ( 2-d \right )\Tilde{\alpha}_{0,1}+\left ( d-2 \right )^{2}\Tilde{\alpha}_{0,2}+\left ( 2-d \right )^{2}\Tilde{\alpha}_{1,1}~,\\
\frac{16\Tilde{f}_{0} }{d^{2}-4} &=4\alpha_{0,0}-2d\alpha_{0,1}+\left ( d^{2}+4 \right )\alpha_{0,2}-\left ( 2-d \right )\left ( 2+d \right )\alpha_{1,1}~,\\
0 &= 4\Tilde{\alpha}_{0,0}-2d\Tilde{\alpha}_{0,1}+\left ( d^{2}+4 \right )\Tilde{\alpha}_{0,2}-\left ( 2-d \right )\left ( 2+d \right )\Tilde{\alpha}_{1,1}~,\\
\frac{4\left ( d+2 \right )f_{0} +2\left(d+6\right)\Tilde{f}_{0} }{\left(d+2\right)^{2}\left(d-2\right)} &=-\alpha_{0,1}+\left ( d+2 \right )\alpha_{0,2}-\left ( 2-d \right )\alpha_{1,1}~,\\
 d-2&=4\alpha_{0,0}-2d\alpha_{0,1}+\left ( d^{2}+4 \right )\alpha_{0,2}-\left ( 2-d \right )\left ( 2+d \right )\alpha_{1,1}~, \\
 -d&=-4\alpha_{0,0}+2d\alpha_{0,1}-\left ( d^{2}+4 \right )\alpha_{0,2}+\left ( 2-d \right )\left ( 2+d \right )\alpha_{1,1}-2\Tilde{\alpha}_{0,1}\\&\quad +2\left ( d+2 \right )\Tilde{\alpha}_{0,2}-2\left ( 2-d \right )\Tilde{\alpha}_{1,1}~.
\end{aligned}
\end{equation}
The solution is
\begin{equation}
\begin{aligned}
\tilde{f}_{0}&=\frac{\left ( d-2 \right )^{2}\left ( d+2 \right )}{16}~,\\
\Tilde{\alpha}_{0,0}&=-\frac{d^{2}-4}{8}+\frac{\left ( d-2 \right )^{2}}{4}\Tilde{\alpha}_{1,1}~,\\
\Tilde{\alpha}_{0,1}&=-\frac{d}{2}+\left ( d-2 \right )\Tilde{\alpha}_{1,1}~,\\ 
\Tilde{\alpha}_{0,2}&=-\frac{1}{2},\\ 
\alpha_{0,0}&=\frac{\left ( d-2 \right )\left ( d+2 \right )^2}{64}+\frac{1}{2}f_{0}+\frac{\left ( d-2 \right )^{2}}{4}\alpha_{1,1}~,\\
\alpha_{0,1}&=\frac{d\left ( d-2 \right )\left ( d+6 \right )}{16\left ( d+2 \right )}+\frac{2d}{d^{2}-4}f_{0}+\left ( d-2 \right )\alpha_{1,1}~,\\
\alpha_{0,2}&=\frac{\left ( d-2 \right )\left ( d+6 \right )}{16\left ( d+2 \right )}+\frac{2}{d^{2}-4}f_{0}~. 
\end{aligned}
\end{equation}

Altogether, then, we are led to the extrinsic counterterm action
\begin{equation}
\begin{aligned}
S\ext &= \int d^{d}x\sqrt{h}\left [ \vphantom{\frac{\left ( d-2 \right )^{2}}{4}}\frac{d-2}{4}\Phi^{2}-\frac{r}{2}\left ( \frac{d^{2}-4}{4} \Phi^{2} +d\,\Phi \dot{\Phi }+\Phi \ddot{\Phi }\right ) \right.\\
 &\quad\quad\quad\quad \quad\quad +\left(\frac{2}{d^{2}-4}f_{0}+\frac{\left(d+6\right)\left(d-2\right)}{16\left(d+2\right)}\right) \left ( \frac{d^{2}-4}{4}\Phi^{2} +d\Phi \dot{\Phi }+\Phi \ddot{\Phi } \right ) \\
 &\quad\quad\quad\quad\quad\quad +\left.\left ( \alpha _{1,1}+\tilde{\alpha }_{1,1}r \right )\left ( \frac{\left ( d-2 \right )^{2}}{4}\Phi ^{2}+\left ( d-2 \right )\Phi \dot{\Phi }+\dot{\Phi }^{2} \right ) \right ].
\end{aligned}
 \label{Sktctlogesque}
\end{equation}
With these counterterms, the on-shell evaluation of the renormalized action is finite, and given by
\begin{equation}
     S\ren=-\int d^{d}x\left ( \phi _{\left ( \Delta \right )}-\frac{\left ( d-2 \right )\left ( d^{2}-20 \right )+32f_{0}}{4\left ( d^{2}-4 \right )}\Tilde{\phi} _{\left ( \Delta \right )} \right )\phi _{\left ( 0 \right )}~.
\end{equation}
We see here that $f_{0}$ is related to the expected renormalization-scheme dependence that arises when $\Delta=\frac{d}{2}+1$ \cite{deHaro:2000vlm}. Meanwhile, all the terms proportional to $\alpha_{1,1}$ and $\tilde{\alpha}_{1,1}$ vanish in the limit $r\rightarrow\infty$. Adjusting these coefficients, we can reexpress the extrinsic counterterm action in a variety of forms that at first sight would appear to be different, without modifying the value of the renormalized action.

The intrinsic counterterms for this conformal dimension are given by \cite{deHaro:2000vlm}
\begin{equation}
\begin{aligned}
S\intr&= \int d^{d}x\sqrt{h}\left [ -\frac{\Delta -d}{2}\Phi ^{2}+\frac{1}{2}\left(r+2c\right)\Phi \Box_{ h}\Phi \right ]\\
 &= \int d^{d}x\sqrt{h}\left [ \frac{d-2}{4}\Phi ^{2}-\frac{r}{2}\left ( \frac{d^{2}-4}{4} \Phi^{2} +d\,\Phi \dot{\Phi }+\Phi \ddot{\Phi }\right )\right.\\
 &\quad \quad\quad\quad\quad\quad\left.-c \left ( \frac{d^{2}-4}{4} \Phi^{2} +d\,\Phi\dot{\Phi }+\Phi\ddot{\Phi }\right ) \right ],
\end{aligned}
 \label{Sctlogesque}
\end{equation}
where in the second equality we have used the equation of motion to rewrite everything in terms of extrinsic derivatives. Clearly (\ref{Sctlogesque}) has the same structure as our extrinsic result (\ref{Sktctlogesque}). The comparison shows that the scheme-dependent coefficients of both formalisms, $f_0$ and $c$, are related through
\begin{equation}
    c=-\left(\frac{2}{d^{2}-4}f_{0}+\frac{\left(d+6\right)\left(d-2\right)}{16\left(d+2\right)}\right)~.
\end{equation}

\section{Conclusions}\label{conclusiones}

In this work, 
we have developed a new, purely \emph{extrinsic} approach to holographic renormalization, and we have exemplified it by determining the counterterms $S\ext$ for a free minimally-coupled scalar field on an asymptotically AdS background, in terms of radial derivatives of the scalar field at the boundary. This new approach satisfies the  all-important requirements of rendering the on-shell action finite and maintaining consistency with the Dirichlet variational principle for the holographic source. The method necessitates no previous knowledge of the standard, intrinsic counterterms $S\intr$. It is reminiscent of the Kounterterm formulation of holographic renormalization in the metric sector \cite{Olea:2005gb,Olea:2006vd}.

As discussed in Section~\ref{Asymptotic}, the compatibility between extrinsic boundary terms and the Dirichlet variational principle on alAdS geometries relies crucially on the fact that  off-shell field configurations, and therefore also variations of the field, are \emph{asymptotically on shell}, up to normalizable order.  In more detail, in the FG expansion of the scalar field, the different non-normalizable coefficients $\phi_{(k<\Delta)}$ would a priori be free independent parameters within the bulk path integral, and therefore part of the boundary conditions that define the quantum ensemble. However, as the counterterms have to be chosen to renormalize the action at the saddle-point configuration, deviations from the equation of motion at non-normalizable level near the conformal boundary result in a divergent action, causing the path integral contribution to vanish.
The equation of motion is thus automatically imposed asymptotically, even away from the saddle point.  For this reason, when examining
the variational principle, one can use the equation of motion to replace extrinsic derivatives of $\delta\Phi$ with intrinsic derivatives, allowing integration by parts, and consequently, consistency with the desired Dirichlet boundary condition.  

Making use of this insight, in
Sections~\ref{masslesssec}-\ref{Extrinsiclog} we put forth a concrete procedure to construct the extrinsic counterterms for a scalar field, expressed only in terms of radial derivatives at the boundary, for the different cases of interest, i.e., massless and massive scalar in even and odd boundary dimensions. This procedure serves as a proof of principle of the fact that, due to the asymptotic form of alAdS spacetimes (in particular, due to the restrictions imposed by the conformal structure), the distinction between intrinsic and extrinsic derivatives of the fields is immaterial, as both are directly connected through the equation of motion in the FG gauge. 

Interestingly, introducing the concept of asymptotically on-shell variations leads naturally to a series of counterterms determined in terms of eigenfunctions of the radial derivative operator. Indeed, the asymptotic structure of alAdS spacetimes is described by the FG expansion, which is a radial expansion around the conformal boundary, whose different coefficients are fixed by the equation of motion. Thus, the asymptotic conformal structure and the dynamics induce naturally an expansion in terms of radial derivatives, whose direct application to the construction of counterterms is presented in this work. This prescription is closely related to the Hamiltonian formalism of \cite{Papadimitriou:2007sj,Papadimitriou:2010as,Papadimitriou:2003is,Papadimitriou:2004ap,Kraus:1999di,Papadimitriou:2005ii}, which relies on the eigenfunction expansion of the dilatation operator. As it was discussed in the beginning of Section \ref{masslesssec}, the nexus between these two prescriptions lies in the kinematic description of alAdS spacetimes provided by the PBH transformation \cite{Imbimbo:1999bj,Penrose:1985bww,Brown:1986nw}. The latter induces a direct relation between radial diffeomorphisms and Weyl transformations (local dilatations) of the boundary metric. As a consequence, the two operators are not linearly independent and the asymptotically on-shell condition provides the map that allows us to translate between the corresponding bases.

From the extrinsic renormalization procedure, we observe a difference between the cases of massive and massless scalars. Indeed, as discussed in Section \ref{masslesssec}, for massless scalars, requiring finiteness of the variation of the action defines a system of equations for the coefficients of the different radial derivative eigenfunctions whose solution gives the renormalized variation. This expression can be directly integrated from the ansatz motivated by the quadratic nature of the scalar action, by replacing $\delta\Phi\to\Phi$ and multiplying by a factor of $1/2$, thus obtaining the renormalized action. The massless scalar case is also the one more closely related to the extrinsic metric \textit{Kounterterms}, and therefore it is interesting that in this case the extrinsic procedure is most efficient. {In more detail, the Kounterterm method, which has a definite topological and geometric origin, can be thought of as a hybrid renormalization prescription which incorporates both intrinsic and extrinsic derivatives of the metric. The analog of that for the scalar field is examined in Appendix~\ref{translationapp}.}

For massive scalars (Sections \ref{Extrinsicgeneric} and \ref{Extrinsiclog}), finding the solution to the system of equations for the renormalized variation does not directly determine the renormalized action. The reason for this is that there are finite contributions coming from linear combinations of the derivative eigenfunctions, which are not fixed by the variational analysis.\footnote{These terms contribute to $\mathcal{F}$ in (\ref{FiniteTerm}).} As discussed in Section \ref{abreviated}, {for $m^2\neq 0$} it is in practice more efficient to simply require the finiteness of the action itself, considering it written as an expansion in the derivative eigenfunctions. In any case, one finds full agreement with the usual intrinsic holographic renormalization counterterms, and the Ward identity of diffeomorphisms is satisfied in the standard way, as explained in Appendix~\ref{wardapp}.

Regarding future work that could come about as natural generalizations of the analysis presented here, the most direct avenue to explore is the extension of the purely extrinsic method to the metric sector and to vector fields \cite{Anastasiou:2024}. In the metric case, it should be possible to reframe the Kounterterms in purely extrinsic terms, {via a translation analogous to the one} discussed in Appendix~\ref{translationapp}. Also, in this work we considered only the {simplest} case of a non-backreacting scalar field in {an asymptotically} AdS background. It is important to understand the extrinsic counterterms construction in more generic boundary setups, and when the backreaction of the matter content on the geometry is allowed.

 \section*{Acknowledgments}
The work of IJA is funded by ANID FONDECYT grants No.~11230419 and~1231133. IJA is grateful to Andrei Parnachev and the School of Mathematics at Trinity College Dublin for their hospitality. The work of GA is funded by ANID FONDECYT grants No.~11240059 and No.~1240043. The work of DA, AG, and SPL was partially supported by CONACYT grant A1-S-22886 and DGAPA-UNAM grant IN116823. DA was additionally supported by a DGAPA-UNAM postdoctoral fellowship.

\appendix
\section{Extrinsic-intrinsic counterterms translation scheme}\label{translationapp}
\subsection{Hybrid version of the extrinsic counterterms}
In this Appendix we will deduce the extrinsic counterterms following a different path, starting with the intrinsic counterterm action and replacing one Laplacian with radial derivatives, via the equation of motion. As before, thanks to the insight gained in Section~\ref{Asymptotic}, we are free to use the equation of motion at the level of the action, without concern of clashing with the variational principle. The equation of motion is \eqref{EOM}:
\begin{equation}
    \frac{1}{ \sqrt{g} } \partial_{ r }\left( \sqrt{g}g^{ rr } \partial_{ r }\Phi  \right)+ \frac{1}{ \sqrt{g} } \partial_{ i }\left( \sqrt{g}h^{ ij } \partial_{ j }\Phi  \right)= m^{2} \Phi~.
\end{equation}
The explicit relation between intrinsic derivatives and the extrinsic (radial) derivative is
\begin{equation}
    \Box_{h} \Phi =m^{2} \Phi -\dot{\Phi} -\ddot{\Phi}~,
    \label{Laplh}
\end{equation}
where we define $\Box_{h} \equiv h^{ ij }\nabla_{  i  }\nabla_{  j  },$ and we remember the relation between the mass of the scalar field and the conformal dimension of the dual operator, given by $m^{2}=\Delta\left(\Delta-d\right).$ 

The intrinsic counterterm action for $\frac{d}{2}+1< \Delta < \frac{d}{2}+2$ is given by \cite{Skenderis:2002wp}
\begin{equation}
S\intr= \int{ d^{d}x} \sqrt{h}\left[ \frac{d- \Delta }{2} \Phi ^{2}+ \frac{1}{2 \left(2 \Delta -d-2\right)} \Phi \Box_{h} \Phi   \right]~.
\label{sct1kt1}
\end{equation}
We use (\ref{Laplh}) to rewrite the above expression in terms of $\Phi$ and its radial derivatives. We obtain that $S\intr$ then takes the form
\begin{equation}
S\intr= \int d^{d}x\,\sqrt{h}\,\left [ \left ( \frac{d-\Delta }{2}+\frac{\Delta \left ( \Delta -d \right )}{2\left ( 2\Delta -d-2 \right )} \right )\Phi ^{2}- \frac{\Phi}{2\left ( 2\Delta -d-2 \right )}\left(d\dot{\Phi}+\ddot{\Phi}\right)\right ]~.
\label{skt-d}
\end{equation}
Notice that this action coincides with {the extrinsic counterterm action $S\ext$ given in} (\ref{Sctextfirstwindow}), which was derived with the radial derivative eigenfunctions method developed in the main text. 

For the next window of possible values of $\Delta$, $\frac{d}{2}+2<\Delta<\frac{d}{2}+3$, the intrinsic counterterm action is \cite{Skenderis:2002wp}
\begin{equation}
    S\intr= \int{ d^{d}x} \sqrt{h}\left[ \frac{d- \Delta }{2} \Phi ^{2}+ \frac{1}{2 \left(2 \Delta -d-2\right)} \Phi \Box_{h} \Phi +\frac{\Phi \left ( \Box_{h} \right )^{2} \Phi}{2\left(2 \Delta -d-2\right)^{2}\left ( 2\Delta -d-4 \right )}  \right].
    \label{kontcont2}
\end{equation}
Via the equation of motion (\ref{Laplh}), this can be reexpressed as  
\begin{equation}
\begin{aligned}
{S\hyb} &= \int d^{d}x\sqrt{h}\left [ \left ( \frac{d-\Delta }{2}+\frac{\Delta\left ( \Delta -d \right )}{2\left ( 2\Delta -d-2 \right )} \right )\Phi ^{2}-\frac{\Phi}{2\left ( 2\Delta -d-2 \right )}\left ( d\dot{\Phi }+\ddot{\Phi } \right ) \right.\\
 &\quad\quad\quad\quad \quad\quad\left. + \frac{\Box_{h}\Phi}{2\left ( 2\Delta -d-2 \right )^{2}\left ( 2\Delta -d-4 \right )}\left ( \Delta \left ( \Delta -d \right )\Phi-\left ( d\dot{\Phi }+\ddot{\Phi } \right ) \right )\right ] .
 \label{skt-3-l}
\end{aligned}
\end{equation}
Notice that when more than one longitudinal Laplacian is present, we {choose in this Appendix to} only utilize the equation of motion once, defining a hybrid scheme {that is intermediate} between {purely} intrinsic and {purely} extrinsic counterterms. {It is for this reason that we have chosen to label this form of the counterterm action with the $^{\mbox{\tiny hyb}}$ superindex.} {The} structure {of (\ref{skt-3-l})} is similar to the Kounterterm action of the metric \cite{Olea:2005gb,Olea:2006vd}, where there is dependence {on} both the intrinsic and extrinsic curvatures. So, for $\frac{d}{2}+n<\Delta<\frac{d}{2}+n+1$ with $n>1$, the longitudinal Laplacians {acting on the scalar field} play an analogous role to the intrinsic curvature in the metric sector, i.e., $ \Box_{h}{\Phi}\leftrightarrow R_{i}^{j}\left [ h \right ]$ \cite{Papadimitriou:2004ap}.\\

{As our final example,} we write the hybrid reformulation of $S\intr$ for the third window of values of $\Delta$, corresponding to $n=3$, obtaining  
\begin{equation}
\begin{aligned}
{S\hyb} &= \int d^{d}x\sqrt{h}\left [\vphantom{\frac{\left(\Box_{h}\right)^{2}\Phi\left ( \Delta \left ( \Delta -d \right )\Phi-\left ( d\dot{\Phi }+\ddot{\Phi } \right ) \right )}{2\left ( 2\Delta -d-2 \right )^{2}\left ( 2\Delta -d-4 \right )\left ( 2\Delta -d-6 \right )}} \left ( \frac{d-\Delta }{2}+\frac{\Delta\left ( \Delta -d \right )}{2\left ( 2\Delta -d-2 \right )} \right )\Phi ^{2}-\frac{\Phi}{2\left ( 2\Delta -d-2 \right )}\left ( d\dot{\Phi }+\ddot{\Phi } \right ) \right.\\
 &\quad\quad\quad\quad \quad\quad + \frac{\Box_{h}\Phi}{2\left ( 2\Delta -d-2 \right )^{2}\left ( 2\Delta -d-4 \right )}\left ( \Delta \left ( \Delta -d \right )\Phi-\left ( d\dot{\Phi }+\ddot{\Phi } \right ) \right ) \\
 &\quad\quad\quad\quad \quad\quad\left. + \frac{\left(\Box_{h}\right)^{2}\Phi\left ( \Delta \left ( \Delta -d \right )\Phi-\left ( d\dot{\Phi }+\ddot{\Phi } \right ) \right )}{2\left ( 2\Delta -d-2 \right )^{2}\left ( 2\Delta -d-4 \right )\left ( 2\Delta -d-6 \right )}\right ].
 \label{sktlapla-4}
\end{aligned}
\end{equation}

As a particular case {of these general results}, we can consider the massless scalar field, which corresponds to $d=\Delta=2n+1$, with $n$ an integer. {Equations} \eqref{skt-d}, \eqref{skt-3-l} and \eqref{sktlapla-4} {then simplify to} 
\begin{subequations}
\begin{equation}
 S\hyb= -\frac{1}{2\left(\Delta-2\right)}\int{ d^{d}x} \sqrt{h} \left(   \Delta\dot{ \Phi} + \ddot{ \Phi} \right)\Phi~,\quad\quad\quad\quad\quad\quad\quad\quad\quad\quad\quad\quad\quad\quad\quad\quad\quad\quad\,
\end{equation}
\begin{equation}
S\hyb=-\frac{1}{2\left(\Delta-2\right)}\int d^{d}x\sqrt{h} \left(   \Delta\dot{ \Phi} + \ddot{ \Phi} \right)\left(\Phi+\frac{\Box_{h}\Phi}{\left(\Delta-2\right)\left(\Delta-4\right)}\right),\quad\quad\quad\quad\quad\quad\quad\quad\quad\quad\quad\quad\quad
\end{equation}
\begin{equation}
S\hyb=-\frac{1}{2\left(\Delta-2\right)}\int d^{d}x\sqrt{h} \left( \Delta\dot{ \Phi} + \ddot{ \Phi} \right)\left[\Phi+\frac{1}{\left(\Delta-2\right)\left(\Delta-4\right)}\left(\Box_{h}\Phi+\frac{2\left(\Box_{h}\right)^{2}\Phi}{\left(\Delta-6\right)}\right)\right],
\end{equation}
\end{subequations}
respectively.

We have obtained the hybrid version of the extrinsic counterterm action for different values of $\Delta$ and $d$. At first sight, this looks different from the $S\ext$ obtained in the main text. However, making use of the equation of motion, we have a real equivalence between them, as in the case of (\ref{Sctextfirstwindow}) and (\ref{skt-d}). This illustrates that, by virtue of the validity of the equation of motion beyond the saddle at the conformal boundary, one can freely exchange radial derivatives by tangential (intrinsic) ones. The hybrid version is functionally more similar to the Kounterterms for the metric case, as they contain both intrinsic Riemannian and extrinsic curvatures of the boundary, as explained above.

\subsection{Extrinsic counterterms vs. hybrid and intrinsic counterterms\label{radialKc}}
In this section we exhibit the explicit relation between the extrinsic counterterms obtained by starting with the radial derivative eigenfunctions method (discussed in the main text) and the hybrid version shown in the previous subsection. To illustrate this idea we focus on the massless scalar field in the second window of possible values of $\Delta$ $\left(n=2\right),$ because in the first window the relation is direct. Remember from (\ref{Sctext2ml}) that the extrinsic counterterm action in this case reads
\begin{equation}
     S\ext=-\frac{1}{2\left ( \Delta -2 \right )\left ( \Delta -4 \right )}\int d^{d}x\sqrt{h}\left [ \Delta \left ( \Delta -6 \right )\dot{\Phi }-6\ddot{\Phi }-\dddot{\Phi } \right ]\Phi ,
\end{equation}
{while} the intrinsic counterterm action (\ref{kontcont2})
{simplifies in the massless case to}
\begin{equation}
    S\intr= \frac{1}{2 \left( \Delta -2\right)}\int{ d^{d}x} \sqrt{h}\left[  \Phi \Box_{h} \Phi +\frac{\Phi \left ( \Box_{h} \right )^{2} \Phi}{\left( \Delta -2\right)\left ( \Delta-4 \right )}  \right].
    \label{sint2massless}
\end{equation}
The equation of motion takes the form
\begin{equation}
    \Box_{h}\Phi=-\Delta\dot{\Phi}-\ddot{\Phi},
    \label{efmmassless2}
\end{equation}
 and substituting \eqref{efmmassless2} in \eqref{sint2massless} we obtain 
\begin{equation}
    {S\hyb}=- \frac{1}{2 \left( \Delta -2\right)}\int{ d^{d}x} \sqrt{h}\left[  \Delta \dot{\Phi }+\ddot{\Phi } +\frac{1}{\left( \Delta -2\right)\left ( \Delta-4 \right )}  \Box_{h}\left ( \Delta \dot{\Phi }+\ddot{\Phi } \right ) \right]\Phi.
    \label{Sintradial2}
\end{equation}
Now, we use the chain rule to expand the Laplacian of the radial derivatives of $\Phi,$ resulting in 
\begin{subequations}
\begin{equation}
 \Box_{h}\partial _{r}\Phi =\partial _{r}\left ( \Box_{h}\Phi  \right )+2\Box_{h}\Phi, \quad\quad\quad\quad\quad
 \label{parbox1}
\end{equation}
\begin{equation}
\,\,\Box_{h}\partial _{r}\partial _{r}\Phi =\partial _{r}\partial _{r}\left ( \Box_{h}\Phi  \right )+4\partial _{r}\left ( \Box_{h}\Phi  \right )+4\Box_{h}\Phi.
\label{parbox2}
\end{equation}
\end{subequations}
Making use of \eqref{parbox1} and \eqref{parbox2} in \eqref{Sintradial2}, we can rewrite (\ref{Sintradial2}) as follows: 
\begin{equation}
\begin{aligned}
 {S\hyb}&= - \frac{1}{2 \left( \Delta -2\right)\left ( \Delta -4 \right )}\int{ d^{d}x} \sqrt{h}\left[ \vphantom{\frac{1}{\Delta -2}} \Delta\left ( \Delta -6 \right ) \dot{\Phi }-6\ddot{\Phi }-\dddot{\Phi }  \right. \\
 &\qquad\qquad\qquad\qquad \left.-\frac{1}{\Delta -2}\left ( { 8\Delta \dot{\Phi }+\left ( 6\Delta +8 \right )\ddot{\Phi }+\left ( \Delta +6 \right )\dddot{\Phi }+\ddddot{\Phi }} \right )  \right ]\Phi~,
\end{aligned}
\end{equation}
where the combination $\sqrt{h}\left(8\Delta \dot{\Phi }+\left ( 6\Delta +8 \right )\ddot{\Phi }+\left ( \Delta +6 \right )\dddot{\Phi }+\ddddot{\Phi }\right)\Phi\rightarrow 0$ asymptotically in the limit $r\rightarrow \infty.$ Therefore the relation between $S\intr$ and $S\ext$ is exhibited, highlighting that an extra part, which is zero at the conformal boundary, has to be added in order to exchange intrinsic quantities for extrinsic ones.   

\section{Ward Identity of diffeomorphisms}
\label{wardapp}
We proceed to illustrate two particular examples of the way in which the bulk correctly captures the Ward identity of diffeomorphisms in the CFT in the presence of an external source $\phi_{(0)}(x)$ for a scalar operator $\mathcal{O}(x)$ \cite{Kalkkinen:2001vg,Bianchi:2001de,Bianchi:2001kw},
\begin{equation}
    \nabla^{j}\left< T_{ij}\right>_{\ren}=-\left< \mathcal{O}\right>_{\ren}\partial _{i}\phi _{\left ( 0 \right )},
\label{ward}
\end{equation}
In~\ref{wardmasslessapp},  we start with a two-dimensional conformal field theory with an operator dual to a massless scalar field $\left(d=\Delta\right)$. In~\ref{wardmassiveapp}, we consider a massive scalar field dual to an operator of dimension $\Delta=\frac{3}{2}.$

\subsection{Massless scalar field}
\label{wardmasslessapp}
{For ease of comparison with the corresponding literature, in this Appendix the} FG expansion for the fields is considered in terms of $\rho=e^{-2r},$ where the induced metric is $h_{ij}(\rho,x)=\frac{1}{\rho}g_{ij}(\rho,x)$, having that
\begin{equation}
    g_{ij}\left(\rho,x\right) =g_{\left ( 0 \right )ij}+\rho g_{\left ( 2 \right )ij}+\cdots+\rho^{\frac{d}{2}} \left ( g_{\left ( d \right )ij}+\log \rho \,\tilde{g}_{\left ( d \right )ij} \right )+\cdots.
\end{equation}
The extrinsic curvature tensor has a FG expansion given by  \cite{Papadimitriou:2004ap}
\begin{equation}
     \begin{aligned}
K_{\left ( 0 \right )ij}\left [ g_{\left ( 0 \right )} \right ] &=g_{\left ( 0 \right )ij}, \\
 K_{\left ( 2 \right )ij}\left [ g_{\left ( 0 \right )} \right ]&= -g_{\left ( 2 \right )ij}\left [ g_{\left ( 0 \right )} \right ],\\
 \vdots \quad\quad & \\
 K_{\left ( d \right )ij}\left [ g_{\left ( 0 \right )} \right ]&= -\frac{d}{2}g_{\left ( d \right )ij}\left [ g_{\left ( 0 \right )} \right ]-\tilde{g}_{\left ( d \right )ij}\left [ g_{\left ( 0 \right )} \right ]+\cdots ,
 \label{Kexpansion}
\end{aligned}
\end{equation}
where $\cdots$ refers to lower order coefficients in the FG expansion of the metric. The stress-energy tensor is given by \cite{Papadimitriou:2004ap}
\begin{equation}
    \left< T_{ij}\right>_{\ren}=-\frac{1}{\kappa ^{2}}\left ( K_{\left ( d \right )ij}-K_{\left ( d \right )}h_{ij} \right ),
\end{equation}
and its covariant derivative reads
\begin{equation}
\nabla^{j}\left< T_{ij}\right>_{\ren}  
 = -\frac{1}{\kappa ^{2}}\left ( \nabla^{j}K_{\left ( d \right )ij}-\nabla_{i}K_{\left ( d \right )} \right )~.
 \label{nablaT}
\end{equation}
For $d=2$ {we know} from (\ref{Kexpansion}) {that} the extrinsic curvature satisfies
\begin{equation}
    K_{\left ( d \right )ij}=-g_{\left ( 2 \right )ij}-\tilde{g}_{\left ( 2 \right )ij}~,
\end{equation}
and its trace is
\begin{equation}
    K_{\left ( d \right )}=-\left ( g_{\left ( 2 \right )}+\tilde{g}_{\left ( 2 \right )} \right )~.
\end{equation}
{Using this in (\ref{nablaT}),}
\begin{equation}
    \nabla^{j}\left<T_{ij} \right>_{\ren}=-\frac{1}{\kappa ^{2}}\left ( - \nabla^{j}g_{\left ( 2 \right )ij}- \nabla^{j}\tilde{g}_{\left ( 2 \right )ij}+\nabla_{i}g_{\left ( 2 \right )}+\nabla_{i}\tilde{g}_{\left ( 2 \right )} \right )~.
    \label{covarianstressenergy}
\end{equation}

Now we consider the Einstein equation in order to rewrite \eqref{covarianstressenergy}, via the Gauss-Codazzi relations \cite{deHaro:2000vlm}, obtaining
\begin{equation}
    \begin{aligned}
 \rho \left [ 2\ddot{g}_{ij}-2\left ( \dot{g}g^{-1}\dot{g} \right )_{ij}+\text{Tr}\left ( g^{-1}\dot{g} \right )\dot{g}_{ij} \right ]&+R_{ij}\left [ g \right ]-\left ( d-2 \right )\dot{g}_{ij}-\text{Tr}\left ( g^{-1}\dot{g} \right )g_{ij} =\\
 &=-\kappa ^{2}\rho ^{d-\Delta -1}\left [ \frac{\Delta \left ( \Delta -d \right )}{d-1}\phi ^{2}g_{ij}+\rho \partial _{i}\phi\partial _{j}\phi  \right ], \\
\nabla_{i}\text{Tr}\left ( g^{-1}\dot{g} \right )- \nabla^{j}\dot{g}_{ij}&= -2\kappa ^{2}\rho ^{d-\Delta -1}\left [ \frac{d-\Delta }{2}\phi \partial _{i}\phi +\rho \dot{\phi }\partial _{i}\phi  \right ],\\
\text{Tr}\left ( g^{-1}\ddot{g} \right )-\frac{1}{2}\text{Tr}\left ( g^{-1}\dot{g}g^{-1}\dot{g} \right ) &=-2\kappa ^{2}\rho ^{d-\Delta -2}\left [ \frac{d\left ( \Delta -d \right )\left ( \Delta -d+1 \right )}{4\left ( d-1 \right )}\phi ^{2} \right. \\
 &\quad\quad\quad\quad\quad\quad\quad \left. -\left ( \Delta -d \right )\rho \phi \dot{\phi }+\rho ^{2}\dot{\phi }^{2}\vphantom{\frac{d\left ( \Delta -d \right )\left ( \Delta -d+1 \right )}{4\left ( d-1 \right )}} \right ].
\end{aligned}
\label{GCrelations}
\end{equation}
The second equation, which considers the $\left(r,x_{i}\right)$ components, is particularly helpful because {it connects with (\ref{covarianstressenergy}). Indeed, upon FG expansion it} gives 
\begin{equation}
    \nabla_{i}g_{\left ( 2 \right )}+\nabla_{i}\tilde{g}_{\left ( 2 \right )}-\nabla^{j}g_{\left ( 2 \right )ij}-\nabla^{j}\tilde{g}_{\left ( 2 \right )ij}=-2\kappa ^{2}\left ( \phi _{\left ( 2 \right )}+\tilde{\phi} _{\left ( 2 \right )} \right )\partial _{i} \phi _{\left ( 0 \right )}~.
    \label{GClogad2}
\end{equation}
{Substituting this} into \eqref{covarianstressenergy}, we obtain 
\begin{equation}
    \nabla^{j}\left< T_{ij}\right>_{\ren}=2\left ( \phi _{\left ( 2 \right )}+\tilde{\phi} _{\left ( 2 \right )} \right )\partial _{i}\phi _{\left ( 0 \right )}~.
\end{equation}
{Since} the expectation value of $\mathcal{O}$ for $d=\Delta=2$ is $\left< \mathcal{O}\right>{\ren}=-2\left ( \phi _{\left ( 2 \right )}+\tilde{\phi} _{\left ( 2 \right )} \right )$ \cite{Skenderis:2002wp}, we have that
\begin{equation}
    \nabla^{j}\left< T_{ij}\right>_{\ren}=-\left< \mathcal{O}\right>{\ren}\,\partial _{i}\phi _{\left ( 0 \right )}~,
\end{equation}
verifying that the Ward identity {(\ref{ward})} is satisfied.
\subsection{Ward identity for a relevant operator}
\label{wardmassiveapp}
Now, we consider a different value of the scaling dimension, namely  
$\Delta=\frac{3}{2},$ {still in $d=2$. In this case, the} FG expansions of the fields change due to the {substantial} backreaction of the scalar field on the metric \cite{Berg:2001ty, Berg:2002hy},
\begin{equation}
     \begin{aligned}
g_{ij} &=g_{\left ( 0 \right )ij}+\rho ^{\frac{1}{2}}g_{\left ( 1 \right )ij}+\rho \left ( g_{\left ( 2 \right )ij}+\log \rho\, \tilde{g}_{\left ( 2 \right )ij} \right ), \\
 \Phi &=\rho ^{-\frac{1}{4}} \left ( \phi _{\left ( 0 \right )}+\rho ^{\frac{1}{2}}\left ( \phi _{\left ( 1 \right )}+\log \rho \,\tilde{\phi } _{\left ( 1 \right )} \right )+\rho\, \phi _{\left ( 2 \right )} \right )~.
\end{aligned}
\end{equation}
{We} compute {separately} the two contributions {to the renormalized stress-energy tensor:} the {regularized} tensor $T\reg ,_{ij}$ {arising from the bare bulk action} and the counterterm {tensor} $T\ct ,_{ij}$. To order $\rho^{0},$ {the former is given by}  
\begin{equation}
    \begin{aligned}
 T\reg,_{ij}&= \vphantom{\frac{2}{\sqrt{h}}\frac{\delta }{\delta h^{ij}}\left ( -\frac{1}{2}\int d^{2}x\sqrt{h}K \right )=}-\frac{1}{\kappa ^{2}}\left ( K_{ij}-Kh_{ij} \right )\\
 &=-\frac{1}{\kappa ^{2}}\left ( -2g_{\left ( 2 \right )ij}- \tilde{g}_{\left ( 2 \right )ij}+\frac{1}{2}g_{\left ( 1 \right )}g_{\left ( 1 \right )ij}+g_{\left ( 2 \right )}g_{\left ( 0 \right )ij}\right. \\
 &\quad\quad\quad\quad \left. +\tilde{g}_{\left ( 2 \right )}g_{\left ( 0 \right )ij}-\frac{1}{2}g_{\left ( 1 \right )}^{lm}g_{\left ( 1 \right )lm}g_{\left ( 0 \right )ij} \right ),
\end{aligned}
\end{equation}
where $g_{\left(k\right)}\equiv\text{Tr}g_{\left(k\right)}.$ On the other hand, the counterterm contribution reads 
\begin{equation}
    \begin{aligned}
T\ct,_{ij} &= -\frac{1}{\kappa ^{2}}h_{ij}\left ( 1-\frac{1}{4}R\log \rho +\frac{\kappa ^{2}}{4}\Phi ^{2}\right )\\
 &= -\frac{1}{\kappa ^{2}}g_{\left ( 2 \right )ij}-\frac{1}{4}\left ( 2 g_{\left ( 0 \right )ij}\phi _{\left ( 0 \right )}\phi _{\left ( 1 \right )}+g_{\left ( 1 \right )ij}\phi _{\left ( 0 \right )}^{2}\right ).
\end{aligned}
\end{equation}
The {renormalized} expectation value is given by the {$\epsilon\rightarrow 0$} limit of $T\sub,_{ij}\equiv T\reg,_{ij}+T\ct,_{ij}$ \cite{Berg:2002hy},
\begin{equation}
    \begin{aligned}
\left<T_{ij} \right>\ren &= \displaystyle \lim_{\epsilon  \to 0}T\sub,_{ij}\\
 &= \frac{1}{\kappa ^{2}}\left ( g_{\left ( 2 \right )ij}+\tilde{g}_{\left ( 2 \right )ij} -\frac{1}{2}g_{\left ( 1 \right )}g_{\left ( 1 \right )ij}-g_{\left ( 2 \right )}g_{\left ( 0 \right )ij}\right.\\
&\quad\quad\quad\left. -\tilde{g}_{\left ( 2 \right )}g_{\left ( 0 \right )ij}+\frac{1}{2}g_{\left ( 1 \right )}^{lm}g_{\left ( 1 \right )lm}g_{\left ( 0 \right )ij} \right ) \\
&\quad -\frac{1}{2}g_{\left ( 0 \right )ij}\phi _{\left ( 0 \right )}\phi _{\left ( 1 \right )}-\frac{1}{4}g_{\left ( 1 \right )ij}\phi _{\left ( 0 \right )}^{2}~.
\end{aligned}
\end{equation}
We take its covariant derivative and make use of the second Gauss-Codazzi relation \eqref{GCrelations} to order $\rho^{0},$ thus obtaining
\begin{equation}
    \begin{aligned}
\nabla^{j}\left<T_{ij} \right>\ren &= \phi _{\left ( 1 \right )}\partial _{i}\phi _{\left ( 0 \right )}+2\tilde{\phi} _{\left ( 1 \right )}\partial _{i}\phi _{\left ( 0 \right )}-\frac{1}{2\kappa ^{2}}g_{\left ( 1 \right )ij} \nabla^{j}g_{\left ( 1 \right )}\\
&\quad -\frac{1}{2\kappa ^{2}}g_{\left ( 1 \right )}\nabla^{j}g_{\left ( 1 \right )ij} -\frac{1}{4}\nabla^{j}g_{\left ( 1 \right )ij}\phi _{\left ( 0 \right )}^{2}-\frac{1}{2}g_{\left ( 1 \right )ij}\phi _{\left ( 0 \right )}\partial ^{j}\phi _{\left ( 0 \right )} .
\end{aligned}
\end{equation}
From the first Gauss-Codazzi relation to order $\rho^{-\frac{1}{2}},$ we have that
\begin{equation}
    \begin{aligned}
g_{\left ( 1 \right )ij} &=-\frac{1}{2}\kappa ^{2}\phi _{\left ( 0 \right )}^{2}g_{\left ( 0 \right )ij}~, \\
 g_{\left ( 1 \right )}&= -\kappa ^{2}\phi _{\left ( 0 \right )}^{2}~,
\end{aligned}
\end{equation}
and therefore 
\begin{equation}
    \nabla^{j}\left<T_{ij} \right>\ren=\phi _{\left ( 1 \right )}\partial _{i}\phi _{\left ( 0 \right )}+2\tilde{\phi} _{\left ( 1 \right )}\partial _{i}\phi _{\left ( 0 \right )}+\frac{1}{4}\kappa^{2}\phi _{\left ( 0 \right )}^{3}\partial_{i}\phi _{\left ( 0 \right )}~.
\end{equation}
From \cite{Berg:2002hy}, the logarithmic term $\tilde{\phi}_{\left(1\right)}$ is given by
\begin{equation}
    \tilde{\phi}_{\left(1\right)}=-\frac{1}{3}\kappa^{2}\phi _{\left ( 0 \right )}^{3}~.
\end{equation}
This term can be neglected{, because} as we have $\Phi$ {only} up to quadratic order in the action. Thus, we can write
\begin{equation}
    \nabla^{j}\left<T_{ij} \right>\ren=\phi _{\left ( 1 \right )}\partial _{i}\phi _{\left ( 0 \right )}~,
\end{equation}
{or in other words,} 
\begin{equation}
    \nabla^{j}\left<T_{ij} \right>_{\ren}=\left ( 2\Delta -d \right )\phi _{\left ( 2\Delta-d  \right )}\partial _{i}\phi _{\left ( 0  \right )}~.
\end{equation}
Using {the fact} that $\left< \mathcal{O}\right>{\ren}=-\left ( 2\Delta -d \right )\phi _{\left ( 2\Delta-d   \right )}$ \cite{Skenderis:2002wp}, we obtain
\begin{equation}
    \nabla^{j}\left<T_{ij} \right>_{\ren}=-\left< \mathcal{O}\right>{\ren}\,\partial _{i}\phi _{\left ( 0  \right )}~,
\end{equation}
showing that the Ward identity {(\ref{ward})} is satisfied. 


\bibliography{Kounterterms}

\providecommand{\href}[2]{#2}\begingroup\raggedright\begin{thebibliography}{10}

\bibitem{Maldacena:1997re}
J.~M. Maldacena, ``{The Large N limit of superconformal field theories and
  supergravity},'' {\em Adv. Theor. Math. Phys.} {\bf 2} (1998) 231--252,
  \href{http://www.arXiv.org/abs/hep-th/9711200}{{\tt hep-th/9711200}}.

\bibitem{Gubser:1998bc}
S.~S. Gubser, I.~R. Klebanov, and A.~M. Polyakov, ``{Gauge theory correlators
  from noncritical string theory},'' {\em Phys. Lett. B} {\bf 428} (1998)
  105--114, \href{http://www.arXiv.org/abs/hep-th/9802109}{{\tt
  hep-th/9802109}}.

\bibitem{Witten:1998qj}
E.~Witten, ``{Anti-de Sitter space and holography},'' {\em Adv. Theor. Math.
  Phys.} {\bf 2} (1998) 253--291,
  \href{http://www.arXiv.org/abs/hep-th/9802150}{{\tt hep-th/9802150}}.

\bibitem{Breitenlohner:1982bm}
P.~Breitenlohner and D.~Z. Freedman, ``{Positive Energy in anti-De Sitter
  Backgrounds and Gauged Extended Supergravity},'' {\em Phys. Lett. B} {\bf
  115} (1982) 197--201.

\bibitem{Breitenlohner:1982jf}
P.~Breitenlohner and D.~Z. Freedman, ``{Stability in Gauged Extended
  Supergravity},'' {\em Annals Phys.} {\bf 144} (1982) 249.

\bibitem{Klebanov:1999tb}
I.~R. Klebanov and E.~Witten, ``{AdS / CFT correspondence and symmetry
  breaking},'' {\em Nucl. Phys. B} {\bf 556} (1999) 89--114,
  \href{http://www.arXiv.org/abs/hep-th/9905104}{{\tt hep-th/9905104}}.

\bibitem{Minces:1999eg}
P.~Minces and V.~O. Rivelles, ``{Scalar field theory in the AdS / CFT
  correspondence revisited},'' {\em Nucl. Phys. B} {\bf 572} (2000) 651--669,
  \href{http://www.arXiv.org/abs/hep-th/9907079}{{\tt hep-th/9907079}}.

\bibitem{Berkooz:2002ug}
M.~Berkooz, A.~Sever, and A.~Shomer, ``{'Double trace' deformations, boundary
  conditions and space-time singularities},'' {\em JHEP} {\bf 05} (2002) 034,
  \href{http://www.arXiv.org/abs/hep-th/0112264}{{\tt hep-th/0112264}}.

\bibitem{Sever:2002fk}
A.~Sever and A.~Shomer, ``{A Note on multitrace deformations and AdS/CFT},''
  {\em JHEP} {\bf 07} (2002) 027,
  \href{http://www.arXiv.org/abs/hep-th/0203168}{{\tt hep-th/0203168}}.

\bibitem{Papadimitriou:2007sj}
I.~Papadimitriou, ``{Multi-Trace Deformations in AdS/CFT: Exploring the Vacuum
  Structure of the Deformed CFT},'' {\em JHEP} {\bf 05} (2007) 075,
  \href{http://www.arXiv.org/abs/hep-th/0703152}{{\tt hep-th/0703152}}.

\bibitem{Susskind:1998dq}
L.~Susskind and E.~Witten, ``{The Holographic bound in anti-de Sitter space},''
  \href{http://www.arXiv.org/abs/hep-th/9805114}{{\tt hep-th/9805114}}.

\bibitem{Peet:1998wn}
A.~W. Peet and J.~Polchinski, ``{UV / IR relations in AdS dynamics},'' {\em
  Phys. Rev. D} {\bf 59} (1999) 065011,
  \href{http://www.arXiv.org/abs/hep-th/9809022}{{\tt hep-th/9809022}}.

\bibitem{Skenderis:1999nb}
K.~Skenderis and S.~N. Solodukhin, ``{Quantum effective action from the AdS /
  CFT correspondence},'' {\em Phys. Lett. B} {\bf 472} (2000) 316--322,
  \href{http://www.arXiv.org/abs/hep-th/9910023}{{\tt hep-th/9910023}}.

\bibitem{deHaro:2000vlm}
S.~de~Haro, S.~N. Solodukhin, and K.~Skenderis, ``{Holographic reconstruction
  of space-time and renormalization in the AdS / CFT correspondence},'' {\em
  Commun. Math. Phys.} {\bf 217} (2001) 595--622,
  \href{http://www.arXiv.org/abs/hep-th/0002230}{{\tt hep-th/0002230}}.

\bibitem{Skenderis:2002wp}
K.~Skenderis, ``{Lecture notes on holographic renormalization},'' {\em Class.
  Quant. Grav.} {\bf 19} (2002) 5849--5876,
  \href{http://www.arXiv.org/abs/hep-th/0209067}{{\tt hep-th/0209067}}.

\bibitem{Fefferman:1985}
C.~Fefferman and C.~R. Graham, ``Conformal invariants,'' in {\em The
  Mathematical Heritage of \'Elie Cartan (Lyon, 1984)}, pp.~95--116.
\newblock Astérisque, 1985.

\bibitem{Papadimitriou:2010as}
I.~Papadimitriou, ``{Holographic renormalization as a canonical
  transformation},'' {\em JHEP} {\bf 11} (2010) 014,
  \href{http://www.arXiv.org/abs/1007.4592}{{\tt 1007.4592}}.

\bibitem{Imbimbo:1999bj}
C.~Imbimbo, A.~Schwimmer, S.~Theisen, and S.~Yankielowicz, ``{Diffeomorphisms
  and holographic anomalies},'' {\em Class. Quant. Grav.} {\bf 17} (2000)
  1129--1138, \href{http://www.arXiv.org/abs/hep-th/9910267}{{\tt
  hep-th/9910267}}.

\bibitem{Papadimitriou:2003is}
I.~Papadimitriou, ``{Holographic renormalization made simple: An example},''
  {\em Subnucl. Ser.} {\bf 41} (2005) 508--514.

\bibitem{Papadimitriou:2004ap}
I.~Papadimitriou and K.~Skenderis, ``{AdS / CFT correspondence and geometry},''
  {\em IRMA Lect. Math. Theor. Phys.} {\bf 8} (2005) 73--101,
  \href{http://www.arXiv.org/abs/hep-th/0404176}{{\tt hep-th/0404176}}.

\bibitem{Kraus:1999di}
P.~Kraus, F.~Larsen, and R.~Siebelink, ``{The gravitational action in
  asymptotically AdS and flat space-times},'' {\em Nucl. Phys. B} {\bf 563}
  (1999) 259--278, \href{http://www.arXiv.org/abs/hep-th/9906127}{{\tt
  hep-th/9906127}}.

\bibitem{Olea:2005gb}
R.~Olea, ``{Mass, angular momentum and thermodynamics in four-dimensional
  Kerr-AdS black holes},'' {\em JHEP} {\bf 06} (2005) 023,
  \href{http://www.arXiv.org/abs/hep-th/0504233}{{\tt hep-th/0504233}}.

\bibitem{Olea:2006vd}
R.~Olea, ``{Regularization of odd-dimensional AdS gravity: Kounterterms},''
  {\em JHEP} {\bf 04} (2007) 073,
  \href{http://www.arXiv.org/abs/hep-th/0610230}{{\tt hep-th/0610230}}.

\bibitem{Kofinas:2007ns}
G.~Kofinas and R.~Olea, ``{Universal regularization prescription for Lovelock
  AdS gravity},'' {\em JHEP} {\bf 11} (2007) 069,
  \href{http://www.arXiv.org/abs/0708.0782}{{\tt 0708.0782}}.

\bibitem{Giribet:2018hck}
G.~Giribet, O.~Miskovic, R.~Olea, and D.~Rivera-Betancour, ``{Energy in
  Higher-Derivative Gravity via Topological Regularization},'' {\em Phys. Rev.
  D} {\bf 98} (2018), no.~4, 044046,
  \href{http://www.arXiv.org/abs/1806.11075}{{\tt 1806.11075}}.

\bibitem{Anastasiou:2020zwc}
G.~Anastasiou, O.~Miskovic, R.~Olea, and I.~Papadimitriou, ``{Counterterms,
  Kounterterms, and the variational problem in AdS gravity},'' {\em JHEP} {\bf
  08} (2020) 061, \href{http://www.arXiv.org/abs/2003.06425}{{\tt 2003.06425}}.

\bibitem{Anastasiou:2021tlv}
G.~Anastasiou, I.~J. Araya, C.~Corral, and R.~Olea, ``{Noether-Wald charges in
  six-dimensional Critical Gravity},'' {\em JHEP} {\bf 07} (2021) 156,
  \href{http://www.arXiv.org/abs/2105.02924}{{\tt 2105.02924}}.

\bibitem{Anastasiou:2023oro}
G.~Anastasiou, I.~J. Araya, C.~Corral, and R.~Olea, ``{Conformal
  Renormalization of topological black holes in AdS$_{6}$},'' {\em JHEP} {\bf
  11} (2023) 036, \href{http://www.arXiv.org/abs/2308.09140}{{\tt 2308.09140}}.

\bibitem{Anastasiou:2022wjq}
G.~Anastasiou, I.~J. Araya, M.~Busnego-Barrientos, C.~Corral, and N.~Merino,
  ``{Conformal renormalization of scalar-tensor theories},''
  \href{http://www.arXiv.org/abs/2212.04364}{{\tt 2212.04364}}.

\bibitem{Miskovic:2008ck}
O.~Miskovic and R.~Olea, ``{Thermodynamics of Einstein-Born-Infeld black holes
  with negative cosmological constant},'' {\em Phys. Rev. D} {\bf 77} (2008)
  124048, \href{http://www.arXiv.org/abs/0802.2081}{{\tt 0802.2081}}.

\bibitem{Miskovic:2010ui}
O.~Miskovic and R.~Olea, ``{Conserved charges for black holes in
  Einstein-Gauss-Bonnet gravity coupled to nonlinear electrodynamics in AdS
  space},'' {\em Phys. Rev. D} {\bf 83} (2011) 024011,
  \href{http://www.arXiv.org/abs/1009.5763}{{\tt 1009.5763}}.

\bibitem{Anastasiou:2019ldc}
G.~Anastasiou, I.~J. Araya, A.~Güijosa, and R.~Olea, ``{Renormalized AdS
  gravity and holographic entanglement entropy of even-dimensional CFTs},''
  {\em JHEP} {\bf 10} (2019) 221,
  \href{http://www.arXiv.org/abs/1908.11447}{{\tt 1908.11447}}.

\bibitem{Balasubramanian:1998sn}
V.~Balasubramanian, P.~Kraus, and A.~E. Lawrence, ``{Bulk versus boundary
  dynamics in anti-de Sitter space-time},'' {\em Phys. Rev. D} {\bf 59} (1999)
  046003, \href{http://www.arXiv.org/abs/hep-th/9805171}{{\tt hep-th/9805171}}.

\bibitem{Skenderis:2008dh}
K.~Skenderis and B.~C. van Rees, ``{Real-time gauge/gravity duality},'' {\em
  Phys. Rev. Lett.} {\bf 101} (2008) 081601,
  \href{http://www.arXiv.org/abs/0805.0150}{{\tt 0805.0150}}.

\bibitem{Skenderis:2008dg}
K.~Skenderis and B.~C. van Rees, ``{Real-time gauge/gravity duality:
  Prescription, Renormalization and Examples},'' {\em JHEP} {\bf 05} (2009)
  085, \href{http://www.arXiv.org/abs/0812.2909}{{\tt 0812.2909}}.

\bibitem{Marolf:2006nd}
D.~Marolf and S.~F. Ross, ``{Boundary Conditions and New Dualities: Vector
  Fields in AdS/CFT},'' {\em JHEP} {\bf 11} (2006) 085,
  \href{http://www.arXiv.org/abs/hep-th/0606113}{{\tt hep-th/0606113}}.

\bibitem{Compere:2008us}
G.~Compere and D.~Marolf, ``{Setting the boundary free in AdS/CFT},'' {\em
  Class. Quant. Grav.} {\bf 25} (2008) 195014,
  \href{http://www.arXiv.org/abs/0805.1902}{{\tt 0805.1902}}.

\bibitem{BrinkHenneaux}
L.~Brink and M.~Henneaux, {\em {Principles of String Theory}}.
\newblock Plenum Press, New York, 1988.

\bibitem{Aros:1999kt}
R.~Aros, M.~Contreras, R.~Olea, R.~Troncoso, and J.~Zanelli, ``{Conserved
  charges for even dimensional asymptotically AdS gravity theories},'' {\em
  Phys. Rev. D} {\bf 62} (2000) 044002,
  \href{http://www.arXiv.org/abs/hep-th/9912045}{{\tt hep-th/9912045}}.

\bibitem{Mora:2004rx}
P.~Mora, R.~Olea, R.~Troncoso, and J.~Zanelli, ``{Vacuum energy in
  odd-dimensional AdS gravity},''
  \href{http://www.arXiv.org/abs/hep-th/0412046}{{\tt hep-th/0412046}}.

\bibitem{Miskovic:2006tm}
O.~Miskovic and R.~Olea, ``{On boundary conditions in three-dimensional AdS
  gravity},'' {\em Phys. Lett. B} {\bf 640} (2006) 101--107,
  \href{http://www.arXiv.org/abs/hep-th/0603092}{{\tt hep-th/0603092}}.

\bibitem{Heemskerk:2010hk}
I.~Heemskerk and J.~Polchinski, ``{Holographic and Wilsonian Renormalization
  Groups},'' {\em JHEP} {\bf 06} (2011) 031,
  \href{http://www.arXiv.org/abs/1010.1264}{{\tt 1010.1264}}.

\bibitem{Faulkner:2010jy}
T.~Faulkner, H.~Liu, and M.~Rangamani, ``{Integrating out geometry: Holographic
  Wilsonian RG and the membrane paradigm},'' {\em JHEP} {\bf 08} (2011) 051,
  \href{http://www.arXiv.org/abs/1010.4036}{{\tt 1010.4036}}.

\bibitem{Balasubramanian:2012hb}
V.~Balasubramanian, M.~Guica, and A.~Lawrence, ``{Holographic Interpretations
  of the Renormalization Group},'' {\em JHEP} {\bf 01} (2013) 115,
  \href{http://www.arXiv.org/abs/1211.1729}{{\tt 1211.1729}}.

\bibitem{Henningson:1998gx}
M.~Henningson and K.~Skenderis, ``{The Holographic Weyl anomaly},'' {\em JHEP}
  {\bf 07} (1998) 023, \href{http://www.arXiv.org/abs/hep-th/9806087}{{\tt
  hep-th/9806087}}.

\bibitem{Henningson:1998ey}
M.~Henningson and K.~Skenderis, ``{Holography and the Weyl anomaly},'' {\em
  Fortsch. Phys.} {\bf 48} (2000) 125--128,
  \href{http://www.arXiv.org/abs/hep-th/9812032}{{\tt hep-th/9812032}}.

\bibitem{vanRees:2011fr}
B.~C. van Rees, ``{Holographic renormalization for irrelevant operators and
  multi-trace counterterms},'' {\em JHEP} {\bf 08} (2011) 093,
  \href{http://www.arXiv.org/abs/1102.2239}{{\tt 1102.2239}}.

\bibitem{Petkou:1999fv}
A.~Petkou and K.~Skenderis, ``{A Nonrenormalization theorem for conformal
  anomalies},'' {\em Nucl. Phys. B} {\bf 561} (1999) 100--116,
  \href{http://www.arXiv.org/abs/hep-th/9906030}{{\tt hep-th/9906030}}.

\bibitem{Papadimitriou:2005ii}
I.~Papadimitriou and K.~Skenderis, ``{Thermodynamics of asymptotically locally
  AdS spacetimes},'' {\em JHEP} {\bf 08} (2005) 004,
  \href{http://www.arXiv.org/abs/hep-th/0505190}{{\tt hep-th/0505190}}.

\bibitem{Papadimitriou:2016yit}
I.~Papadimitriou, ``{Lectures on Holographic Renormalization},'' {\em Springer
  Proc. Phys.} {\bf 176} (2016) 131--181.

\bibitem{Balasubramanian:1999re}
V.~Balasubramanian and P.~Kraus, ``{A Stress tensor for Anti-de Sitter
  gravity},'' {\em Commun. Math. Phys.} {\bf 208} (1999) 413--428,
  \href{http://www.arXiv.org/abs/hep-th/9902121}{{\tt hep-th/9902121}}.

\bibitem{Caceres:2023gfa}
N.~Caceres, C.~Corral, F.~Diaz, and R.~Olea, ``{Holographic renormalization of
  Horndeski gravity},'' {\em JHEP} {\bf 05} (2024) 125,
  \href{http://www.arXiv.org/abs/2311.04054}{{\tt 2311.04054}}.

\bibitem{Penrose:1985bww}
R.~Penrose and W.~Rindler, {\em {Spinors and Space-Time}}.
\newblock Cambridge Monographs on Mathematical Physics. Cambridge Univ. Press,
  Cambridge, UK, 4, 2011.

\bibitem{Brown:1986nw}
J.~D. Brown and M.~Henneaux, ``{Central Charges in the Canonical Realization of
  Asymptotic Symmetries: An Example from Three-Dimensional Gravity},'' {\em
  Commun. Math. Phys.} {\bf 104} (1986) 207--226.

\bibitem{Anastasiou:2024}
G.~Anastasiou, I.~J. Araya, A.~Güijosa, and S.~Patiño-López , in progress.

\bibitem{Kalkkinen:2001vg}
J.~Kalkkinen, D.~Martelli, and W.~Mueck, ``{Holographic renormalization and
  anomalies},'' {\em JHEP} {\bf 04} (2001) 036,
  \href{http://www.arXiv.org/abs/hep-th/0103111}{{\tt hep-th/0103111}}.

\bibitem{Bianchi:2001de}
M.~Bianchi, D.~Z. Freedman, and K.~Skenderis, ``{How to go with an RG flow},''
  {\em JHEP} {\bf 08} (2001) 041,
  \href{http://www.arXiv.org/abs/hep-th/0105276}{{\tt hep-th/0105276}}.

\bibitem{Bianchi:2001kw}
M.~Bianchi, D.~Z. Freedman, and K.~Skenderis, ``{Holographic
  renormalization},'' {\em Nucl. Phys. B} {\bf 631} (2002) 159--194,
  \href{http://www.arXiv.org/abs/hep-th/0112119}{{\tt hep-th/0112119}}.

\bibitem{Berg:2001ty}
M.~Berg and H.~Samtleben, ``{An Exact holographic RG flow between 2-d conformal
  fixed points},'' {\em JHEP} {\bf 05} (2002) 006,
  \href{http://www.arXiv.org/abs/hep-th/0112154}{{\tt hep-th/0112154}}.

\bibitem{Berg:2002hy}
M.~Berg and H.~Samtleben, ``{Holographic correlators in a flow to a fixed
  point},'' {\em JHEP} {\bf 12} (2002) 070,
  \href{http://www.arXiv.org/abs/hep-th/0209191}{{\tt hep-th/0209191}}.

\end{thebibliography}\endgroup

\bibliographystyle{./utphys}

\end{document}